\documentclass[preprint,3p,sort&compress]{elsarticle}
\usepackage{color}
\usepackage{amsfonts}
\usepackage{amsmath}
\usepackage{amssymb}
\usepackage{subfigure}
\usepackage{graphicx}
\usepackage{tabularx}%
\usepackage{url}
\newcommand{\alf}{\alpha}
\newcommand{\bet}{\beta}
\newcommand{\gam}{\gamma}
\newcommand{\<}{\langle}
\renewcommand{\>}{\rangle}
\newcommand*\e[1]{{e_{ij}^{#1}}}
\newcommand*\mb[1]{\mathbf{#1}}
\newcommand*\mcal[1]{\mathcal{#1}}
\newcommand{\dd}{\mbox{d}}

\newcommand{\hei}{0.24} 

\begin{document}
\title{Stiffest Elastic Networks}
\author{G\'{e}rald Gurtner\fnref{fn1}}
\fntext[fn1]{Present address: Scuola Normale Superiore di Pisa, Piazza dei Cavalieri 7, 56126 Pisa, Italy.}
\author{Marc Durand\corref{cor1}}
\ead{marc.durand@univ-paris-diderot.fr}
\cortext[cor1]{Corresponding author}
\address{Mati\`{e}re et Syst\`{e}mes Complexes (MSC), UMR 7057 CNRS \& Universit\'{e}
Paris Diderot, 10 rue Alice Domon et L\'{e}onie Duquet, 75205 Paris Cedex 13, France}
\date{\today}
\begin{abstract}
The rigidity of a network of elastic beams crucially depends on the specific details of its structure. We show both numerically and theoretically that there is a class of isotropic networks which are stiffer than any other isotropic network with same density. The elastic moduli of these \textit{stiffest elastic networks} are explicitly given. They constitute upper-bounds which compete or improve the well-known Hashin-Shtrikman bounds. We provide a convenient set of criteria (necessary and sufficient conditions) to identify these networks, and show that their displacement field under uniform loading conditions is affine down to the microscopic scale. Finally, examples of such networks with periodic arrangement are presented, in both two and three dimensions.
\end{abstract}

\maketitle

\section{Introduction}
Various elastic systems can be understood as networks of interconnected beams.  
Examples include polymer gels, protein networks and cytoskeletal
structures \cite{Thorpe,Head,Wilhelm,Heussinger,Buxton}, or wood and
bones \cite{Gibson,Ashby,Deshpande,Roberts,Christensen}.
On a length scale much larger than the typical beam length (``macroscopic" scale), such a network can be viewed as a continuous and homogeneous medium characterized by spatially constant elastic moduli.
When this medium is subjected to uniform stresses at its boundaries, it will undergo a homogeneous deformation with a constant strain \cite{Landau}. Such homogeneous deformations are called
\textit{affine}. This picture of affine strain is generally valid at length scales large compared to any characteristic inhomogeneities: the macroscopic displacement field $\bar{\mathbf{u}}(\mathbf{r})$, defined as the spatial average of displacements over a sufficiently large domain surrounding every point $\mathbf{r}$, is affine \cite{DiDonna}.
On a microscopic level however, the displacement field is generally not affine, and beams can deform by a combination of stretching, bending and twisting mechanisms. In addition, the nature of the junctions between beams plays an important role in the stability and mechanical properties of such systems: networks with
junctions that either fix the relative orientation of the
beams (``rigid junctions") or allow free rotation (``free
hinges") can have very different mechanical responses \cite{Wilhelm}.
Actually, junctions in most of the real systems have an intermediate behaviour: a small but finite energy cost is associated with the change of their geometry.

The relationship between the mechanical properties of elastic networks on a macroscopic level and the details of their structures is still poorly understood.
The stiffness of such a system clearly depends on its density $\phi$, defined as the volume of beams per unit volume of material. But for a given value of $\phi$, it is also dramatically affected by the specific spatial arrangement of the elastic phase within the material.
%
On dimensional grounds \cite{Ashby}, the volumetric density of strain energy $\varepsilon$ associated with a stretch-dominated deformation varies linearly with $\phi$, while it scales as $\phi^{2}$ for the deformation of a three-dimensional network dominated by the beam bending mode. 
Thus, for the low-density materials considered here ($\phi\ll1$), a structure deforming primarily through the beam stretching mode is usually much stiffer.
However, the constant of proportionality between $\varepsilon$ and $\phi$ still varies significantly among stretched-dominated networks.

In this paper, we reveal both numerically and theoretically the existence of a class of elastic networks which are stiffer than any other ones having the same symmetry, same density and same elastic phase. These \textit{stiffest networks} deform through the stretching mode exclusively, and their displacement field is affine down to microscopic scale \textit{for any loading conditions}. 
 We derive and analyse the necessary and sufficient conditions under which a network belongs to this class.
We will limit our study to isotropic structures, though the reasoning can be transposed without difficulty to materials with lower symmetries.

In section \ref{Simulations} we investigate numerically the role of node connectivity and structural disorder on the affinity of the displacement field and the stiffness of two-dimensional elastic networks. Our simulations show that stretch-dominated networks with same node connectivity can present significantly different elastic moduli and displacement fields.
In section \ref{Theory} we propose a theoretical framework to rationalize these observations: we derive bounds on the elastic moduli of isotropic networks using a variational approach, and establish simple rules on the geometrical and topological arrangement of beams in a network to make it stiffer for the same amount of material. In section \ref{Analysis}, we analyse the restrictions imposed by these structural conditions, and show that they rationalize our numerical findings. In particular, they prove that stiffest networks deform affinely down to the microscopic scale. Finally, in section \ref{Examples}, we provide examples of both two-dimensional (2d) and three-dimensional (3d) regular structures with highest elastic moduli.

\section{Simulations \label{Simulations}}

\subsection{method}
In order to investigate the interplay between node connectivity, affinity of the strain, and stiffness of the network, we simulated the mechanical behaviour of different 2d regular networks, made of straight and uniform beams (same cross-section and material).  
Five different networks have been simulated in order to inspect the effect of both connectivity $z$ and disorder on the mechanical properties. Three regular ones: hexagonal (connectivity $z=3$), kagome ($z=4$), triangular ($z=6$), and two disordered ones: Voronoi ($z=3$) and Delaunay ($\bar{z}=6$) networks, which are generated from uniformly distributed points on the plane. 

The hamiltonian of a network of beams has three different contributions:  
\begin{equation}
\mathcal{H}=\sum_{(i,j)}\mathcal{H}^{s}_{ij}+\sum_{(i,j)}\mathcal{H}^{b}_{ij}+\sum_i \mathcal{H}^{n}_{i}.
\label{Hamiltonian}
\end{equation}
The first two sums are carried over all beams $(i,j)$ of the network, and represent the contributions of the stretching and bending energy of the beams, respectively:  $\mathcal{H}^{s}_{ij}=\int_0^{l_{ij}}(\kappa_s/2) ( \partial u_s/\partial l)^2 \mathrm{d}l$, and  $\mathcal{H}^{b}_{ij}=\int_0^{l_{ij}}  (\kappa_{b}/2) ( \partial^2 u_{n}/\partial l^2)^2 \mathrm{d}l$, where $u_s$ and $u_n$ are the tangential and normal components of the displacement $\mathbf{u}$, respectively. $\kappa_s$ and $\kappa_b$ are the stretching and bending moduli, related to the Young's modulus of the material $E_0$, the section of the beam $s$ and the second moment of area $I$ by: $\kappa_s = E_0 s$ and $\kappa_b=E_0 I$. In the simulations each network is represented by the set of mobile nodes $\lbrace \mathbf{x}_i \rbrace$ consisting of all beam junctions and midpoints between junctions (the latter so as to include the first bending mode of the beams), with each contribution linearized with respect to changes in the $\lbrace \mathbf{x}_i \rbrace$. The discretized version of the stretching and bending contributions are: $\mathcal{H}_{ij}^s=\kappa_s  \left((\mathbf{u}_{ik}\cdot \mathbf{e}_{ij} )^2 + (\mathbf{u}_{jk}  \cdot \mathbf{e}_{ij})^2 \right)/l_{ij}$, and $\mathcal{H}_{ij}^b=4 \kappa_b  \left((\mathbf{u}_{ik}+\mathbf{u}_{jk}) \times \mathbf{e}_{ij} \right)^2/l_{ij}^3 $, where $\mathbf{u}_{ij}=\mathbf{u}_{j}-\mathbf{u}_{i}$.

\begin{figure}[htbp]
\begin{center}
\includegraphics[height=0.2\textwidth]{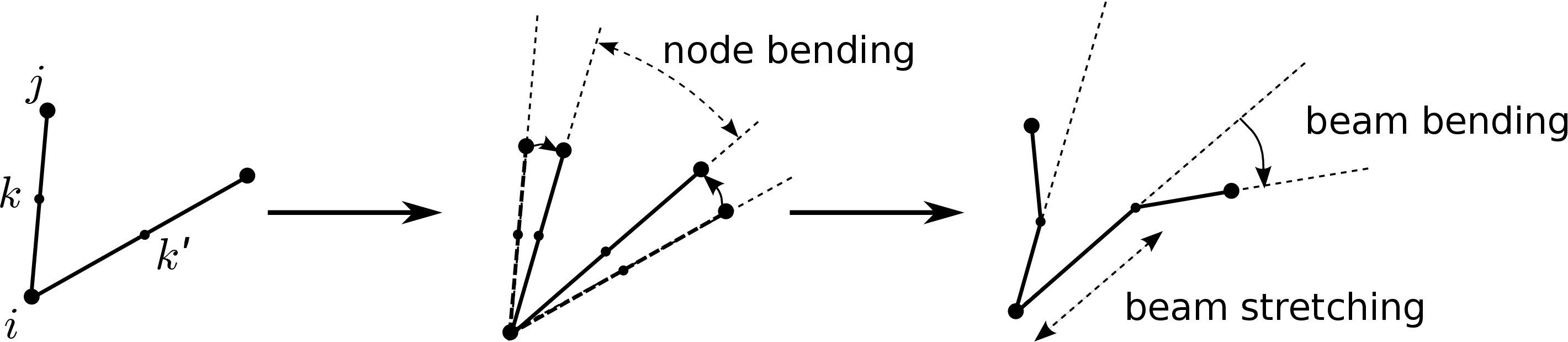}
\label{discrete_beams}
\caption{sketch of discretized beams. Each beam is composed of two segments of equal length. A beam bending energy is associated with a change of angle between the two segments of a beam, a node bending energy is associated with a change of angle between adjoining beams, and a stretching energy is associated with a change of length of the beam segments.}
\end{center}
\end{figure}

The third sum in Eq. (\ref{Hamiltonian}) is carried over all junctions $i$, and $\mathcal{H}^{n}_{i}$ denotes the energy cost associated with the change of geometry of junction $i$. In absence of this energy cost (free hinges), torques cannot be sustained and beams deform through the stretching mode exclusively and act as hookean springs. Conversely, if some ``node rigidity'' is included, other modes of deformation are solicited. 
Different expressions can be considered for $\mathcal{H}^{n}_{i}$ \cite{Head,Wilhelm}. In the simulations presented here, we use $\mathcal{H}^{n}_{i}=\kappa_n \sum_{k,k^\prime\in \mathcal{N}(i)}(\Delta \theta_{ikk^\prime})^2/(l_{ik}+l_{ik^\prime})$, where $\mathcal{N}(i)$ denotes the set of nodes that are connected to node $i$ and $\Delta \theta_{ikk^\prime}$ is the change of the angle between the adjoining beams $(i,k)$ and $(i,k^\prime)$ (see Fig. \ref{discrete_beams}). Because of the analogy of this term with the bending energy of a beam, we will refer this mode of deformation to as ``node bending'' in the following, and $\kappa_n$ as the ``bending modulus of a junction''.

The size of the box $L$ is approximately $40 $ times larger than the average beam length $l$. Range of values used in our simulations for  $\kappa_s$, $\kappa_b$, and $\kappa_n$ are such that $\kappa_s l^2 \gg\kappa_b \gg \kappa_n$ (typically, $\kappa_s l^2 \sim 10^3\kappa_b$, and $\kappa_b \sim 10^3\kappa_n$).
 
%

Either a shear or uniaxial strain $\gamma$ is applied across the top and bottom boundaries, while periodic boundary conditions are used in the other direction. Within our linearized scheme, $\mathcal{H}(\lbrace \mathbf{x}_i \rbrace)$ is a high dimensional paraboloid with a unique global minimum, corresponding to the state of mechanical equilibrium. Hence, the linear system can be solved directly using a \textit{LU} decomposition method included in the UMFPACK routines \cite{umfpack}. 
Once the equilibrium positions are obtained, the associate strain energy can be calculated. Finally the shear modulus $\mu$ and longitudinal modulus $M$ are obtained by equating the energy with $\mu\gamma^2L^2/2$ for shear strain, and $M\gamma^2L^2/2$ for uniaxial strain. Any other elastic moduli can then be calculated. For instance, the Young's modulus $E$ is related to $\mu$ and $M$ as: $E=2\mu(dM-2(d-1)\mu)/((d-1)M-2(d-2)\mu)$ \cite{noteaboutM}. 

\begin{figure}[h]
\begin{center}
\subfigure[]{
\includegraphics[height=\hei\textwidth]{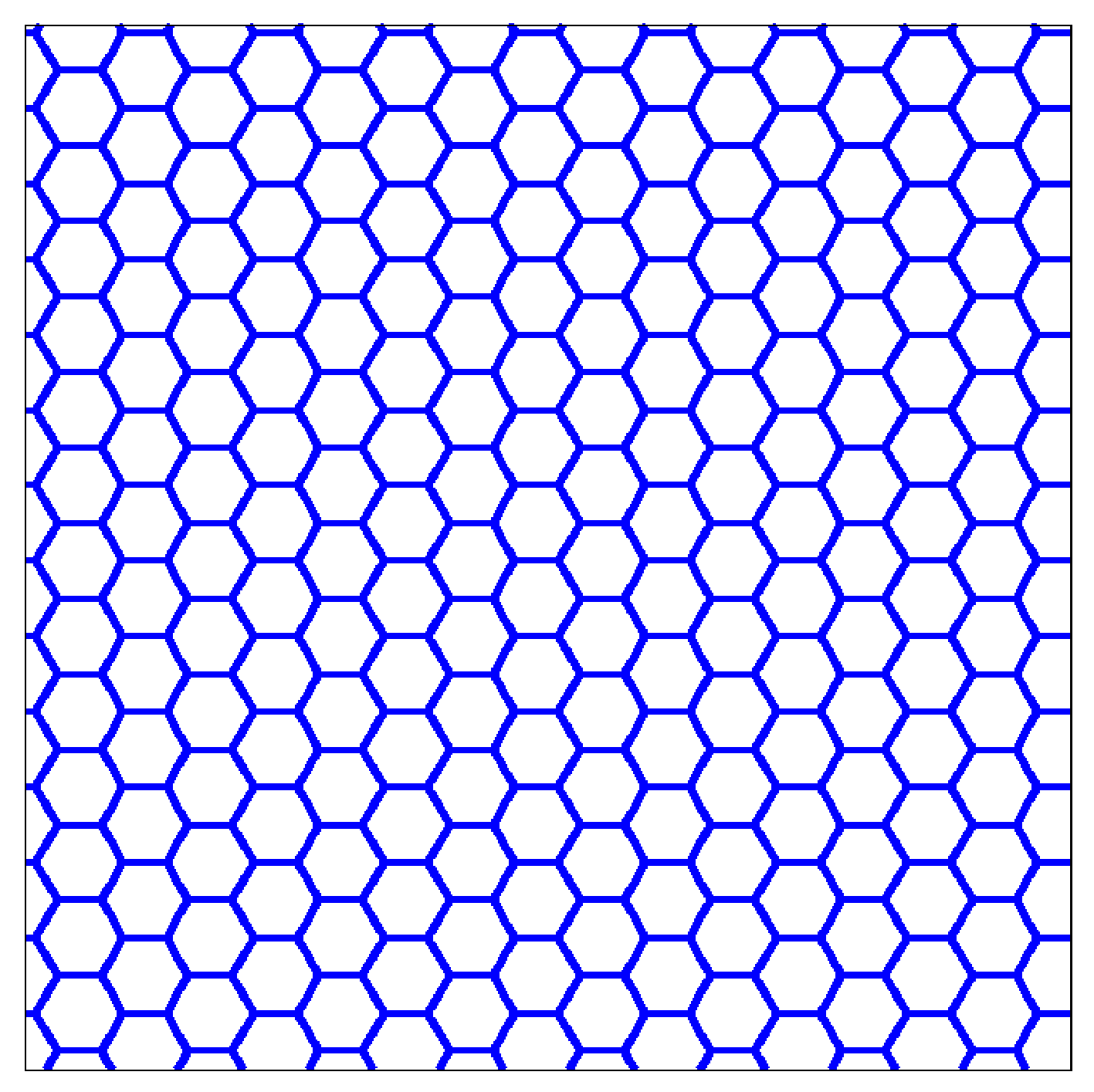}
\label{Hexagonal_poutres}
}
\subfigure[]{
\includegraphics[height=\hei\textwidth]{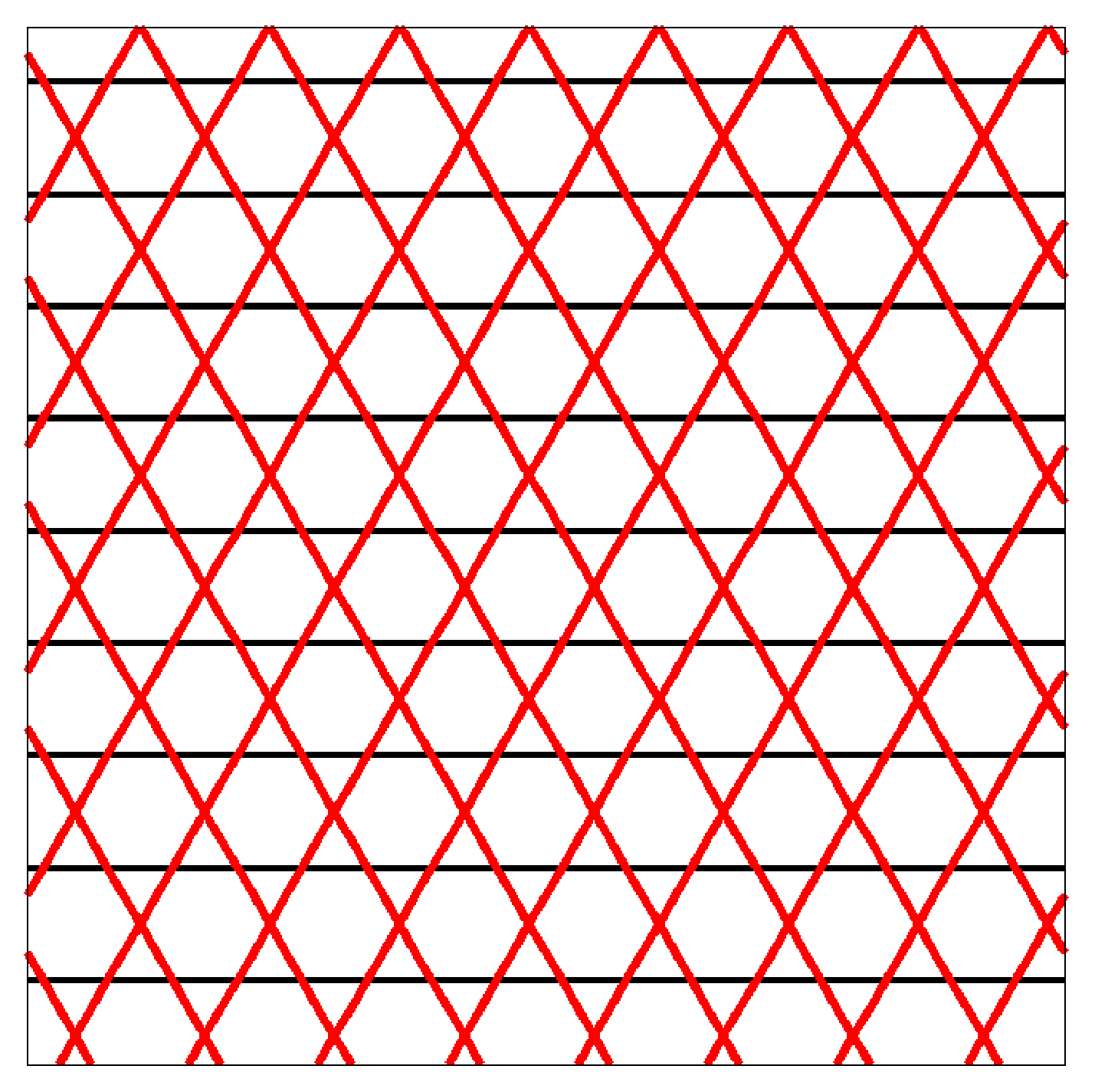}
\label{Kagome_poutres}
}
\subfigure[]{
\includegraphics[height=\hei\textwidth]{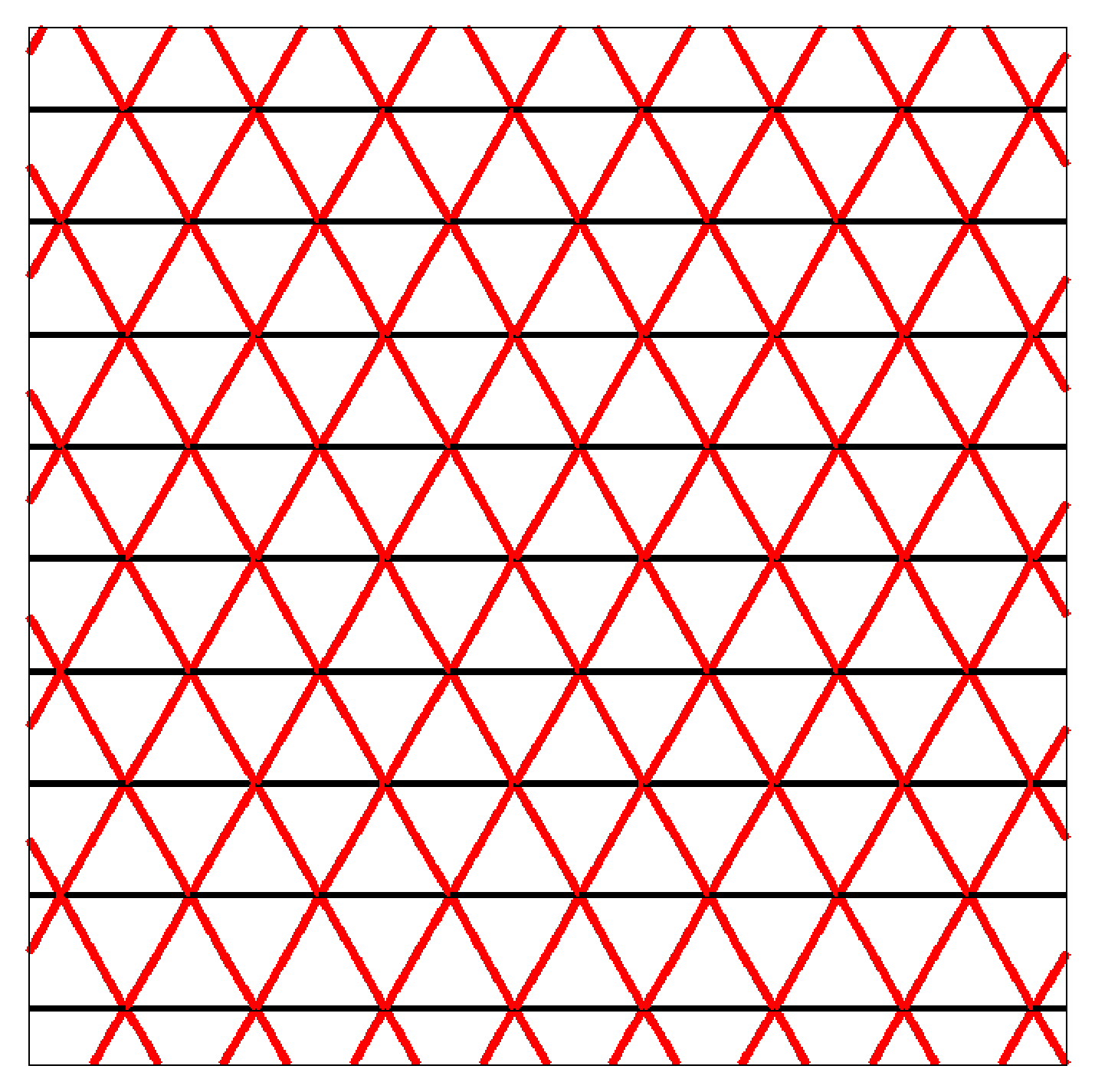}
\label{Triangulaire_poutres}
}
\subfigure[]{
\includegraphics[height=\hei\textwidth]{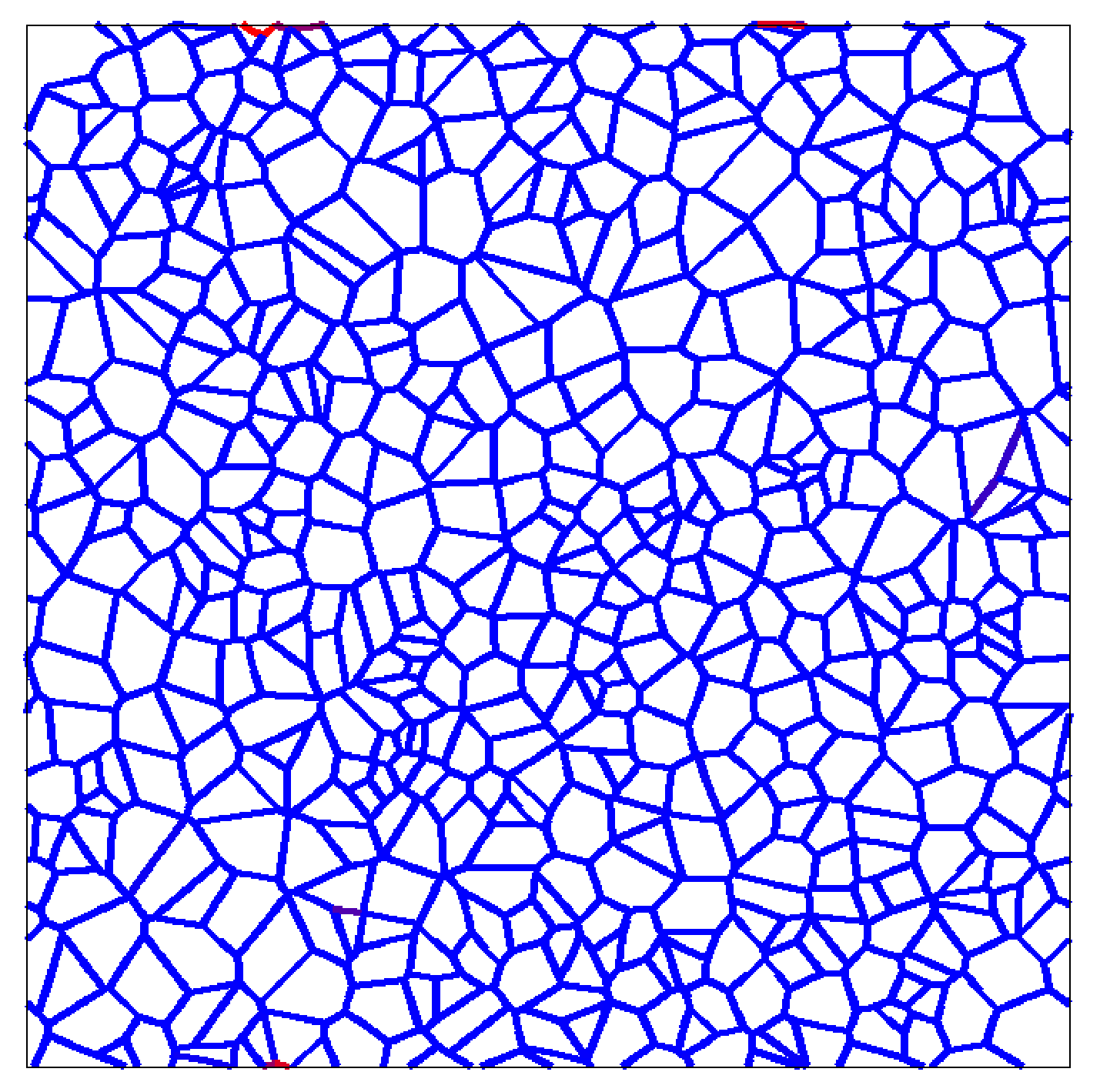}
\label{Voronoi_poutres}
}
\subfigure[]{
\includegraphics[height=\hei\textwidth]{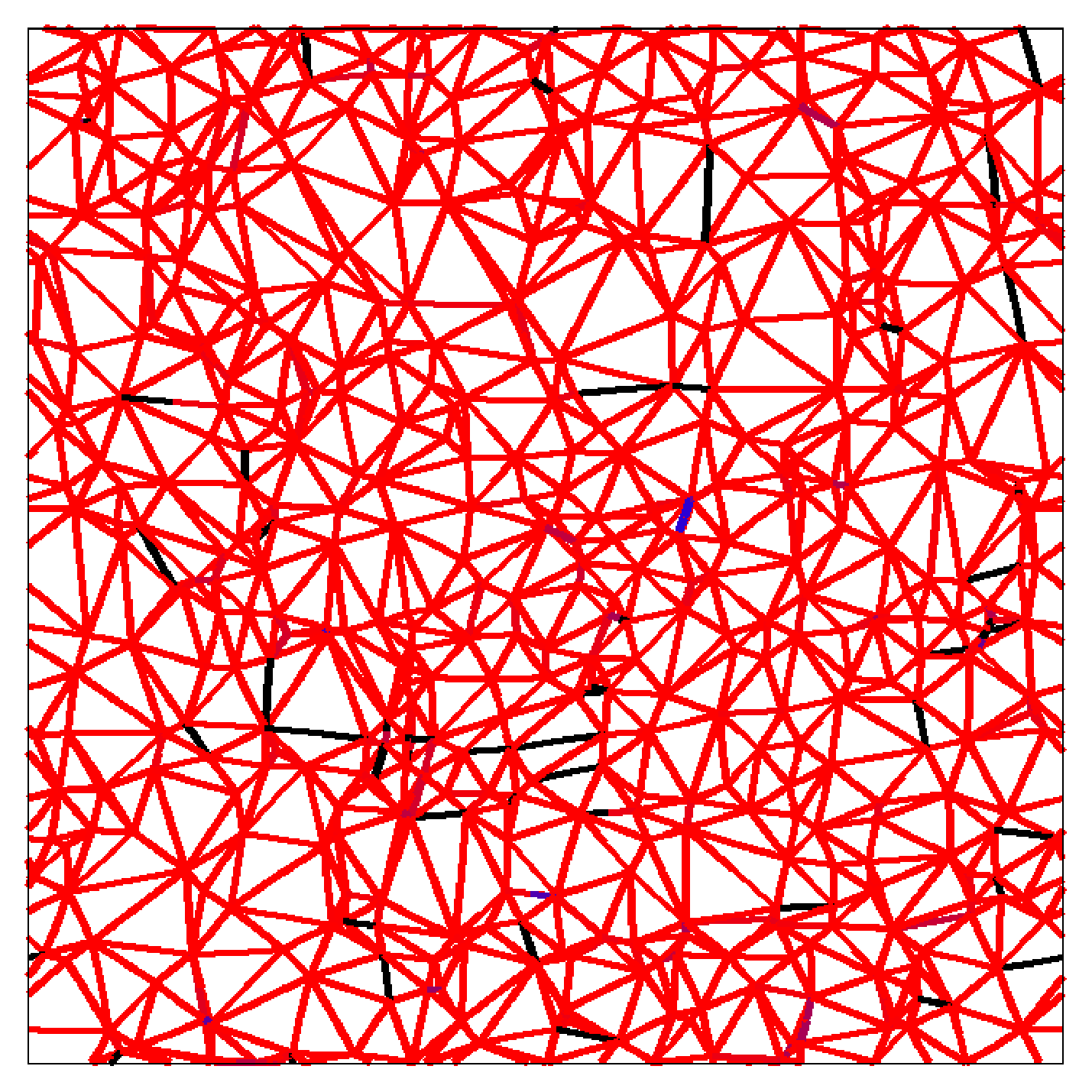}
\label{Delaunay_poutres}
}
\subfigure[]{
\includegraphics[height=0.215\textwidth]{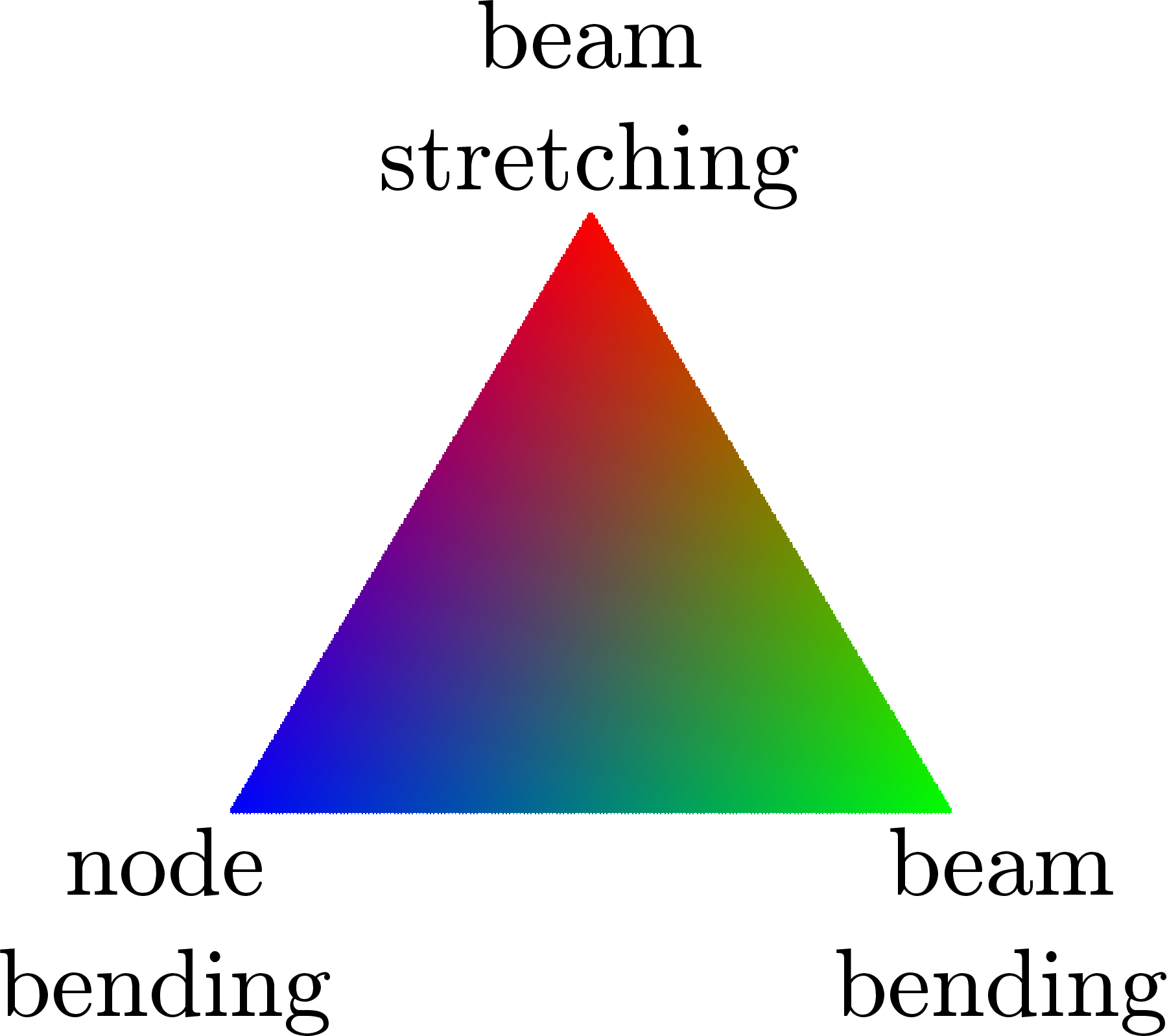}
\label{couleurs}
}
\end{center}
\caption{Repartition of the strain energy in various 2d isotropic networks subject to  horizontal shear deformation: (a) hexagonal network; (b) kagome network; (c) triangular network; (d) Voronoi network; (e) Delaunay network; (f) the calibration triangle  shows what proportion of the total energy is due to beam stretching, beam bending and node bending. Beams are coloured black if relatively undeformed, red if the deformation energy is predominantly beam stretching, green if the deformation energy is predominantly beam bending, and blue if the deformation energy is predominantly node bending of the two associated junctions.}
\label{repartition_energie}
\end{figure}

\subsection{Results}

For each simulated network, we represent the repartition of energy between beam stretching, beam bending, and node bending modes, and measure the affinity of the displacement field. Different parameters have been proposed in literature \cite{Head,Buxton,Liu} to quantify the degree of (non)affinity of a displacement field, but most of them are global measures of the affinity. We use instead the local affinity measure $m_i$ defined at every node (junction and midpoint) $i$ as $m_i^2 = (\mathbf{\bar{u}}_i-\mathbf{u}_i))^2/(\gamma L)^2 $, where $\mathbf{\bar{u}}_i$ is the macroscopic, affine, displacement field of the node $i$. 

Figure \ref{repartition_energie} shows the repartition of energy between beam stretching, beam bending and node bending when the networks are subjected to shear strain (similar results, not shown here, are obtained for uniaxial strain). Hexagonal and Voronoi networks deform mainly through the node bending mode, while kagome, triangular and Delaunay networks deform mainly through the stretching mode. These results are consistent with Maxwell criterion on rigidity of 2d frameworks \cite{Deshpande, Dunlop, Thorpe3, Heussinger2}: networks with connectivity above $4$ are rigid; they deform primarily through the stretching of beams, while those with connectivity below $4$ are soft; that is, they deform through floppy modes if $\kappa_n=0$ \cite{Kellomaki} (free hinges), or through the bending of beams and nodes if $\kappa_n \neq 0$. The repartition of energy between the beam and node bending modes depends on the relative importance of $\kappa_b$ and $\kappa_n$ (rigid junctions with fixed angles correspond to the limit $\kappa_n \gg \kappa_b$). 
In our simulations the bending of beams is quasi-absent (see Fig. \ref{repartition_energie}), because $\kappa_n \ll \kappa_b$.

\begin{figure}[h]
\begin{center}
\subfigure[]{
\includegraphics[height=\hei\textwidth]{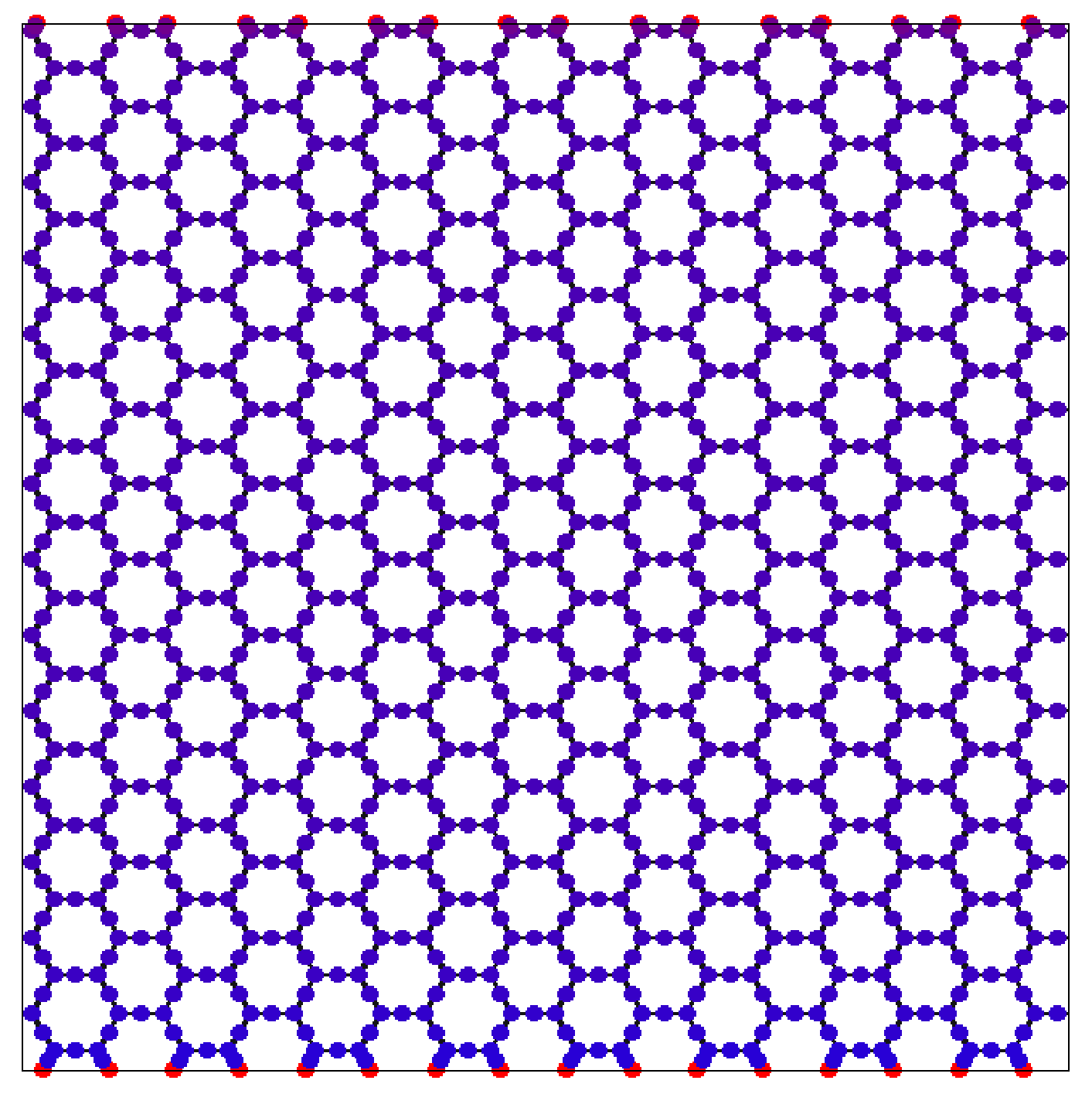}
\label{Hexagonal_noeuds}
}
\subfigure[]{
\includegraphics[height=\hei\textwidth]{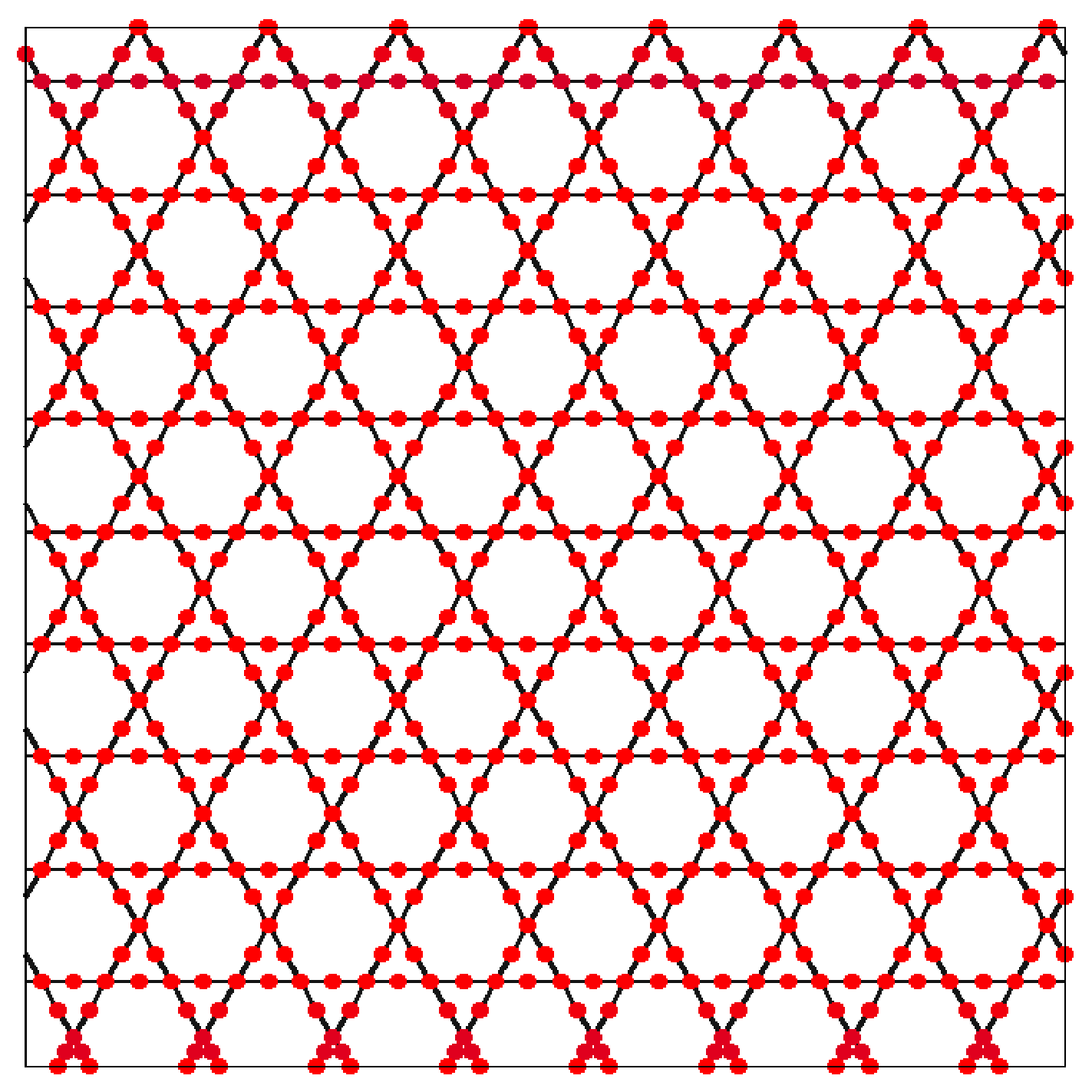}
\label{Kagome_noeuds}
}
\subfigure[]{
\includegraphics[height=\hei\textwidth]{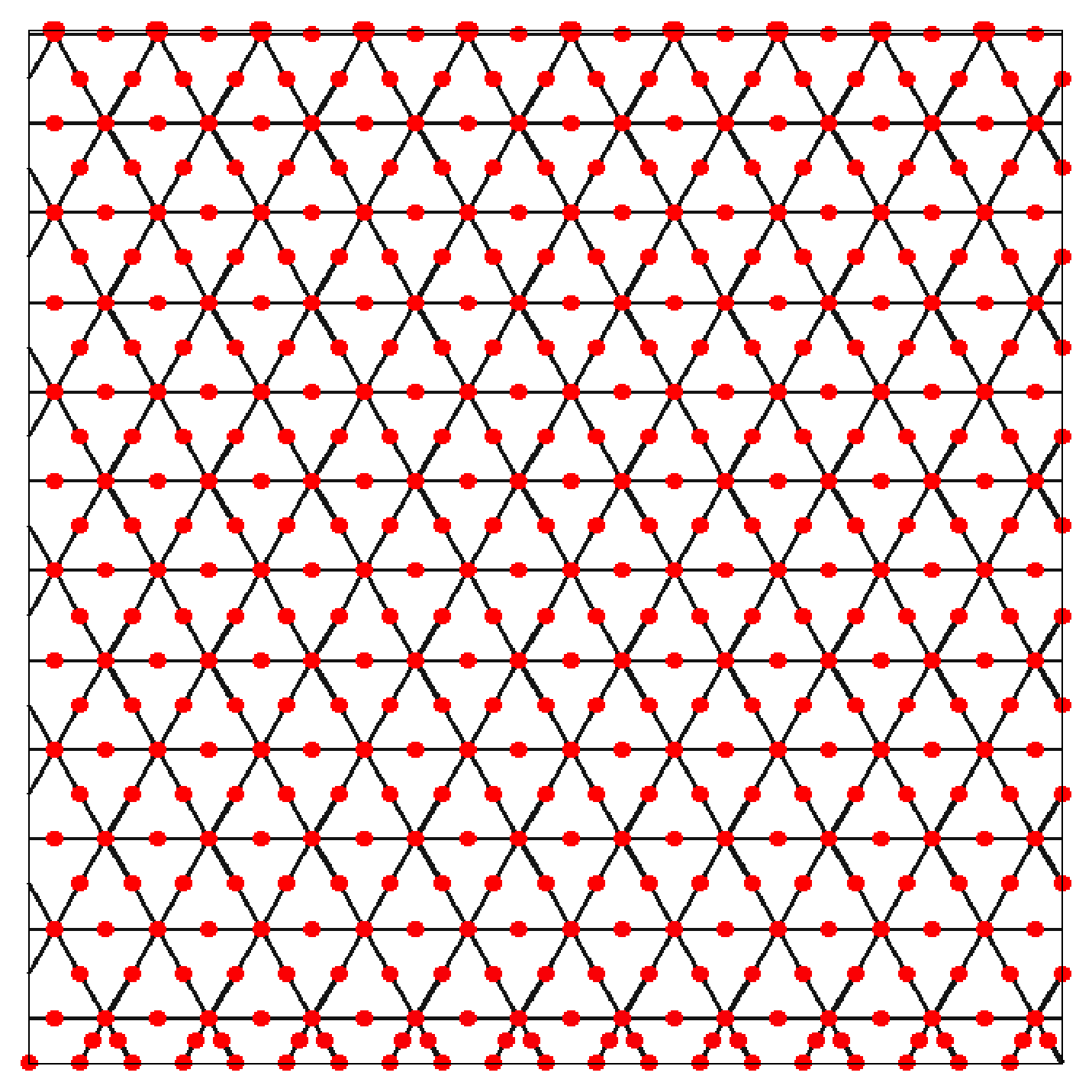}
\label{Triangulaire_noeuds}
}
\subfigure[]{
\includegraphics[height=\hei\textwidth]{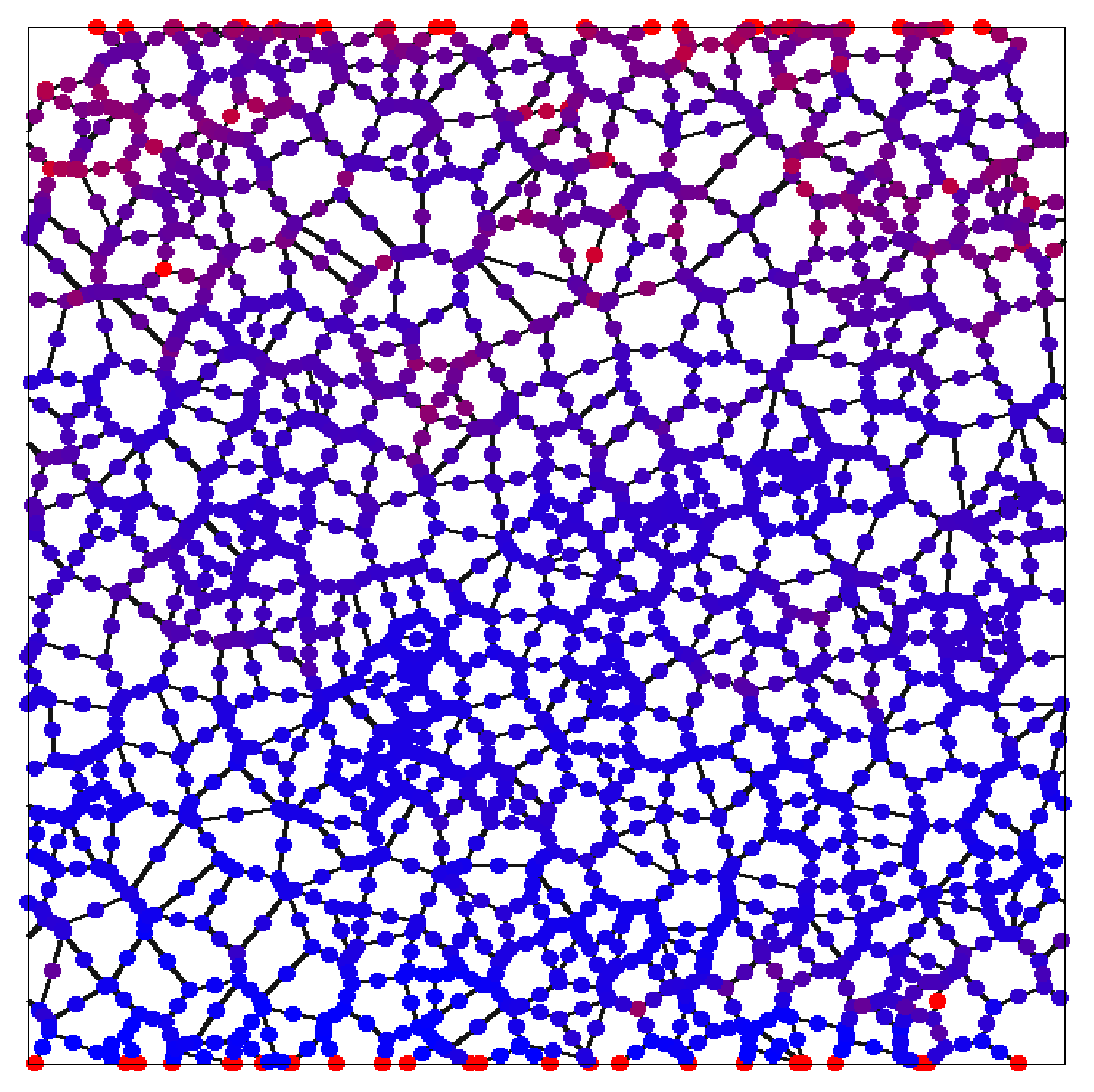}
\label{Voronoi_noeuds}
}
\subfigure[]{
\includegraphics[height=\hei\textwidth]{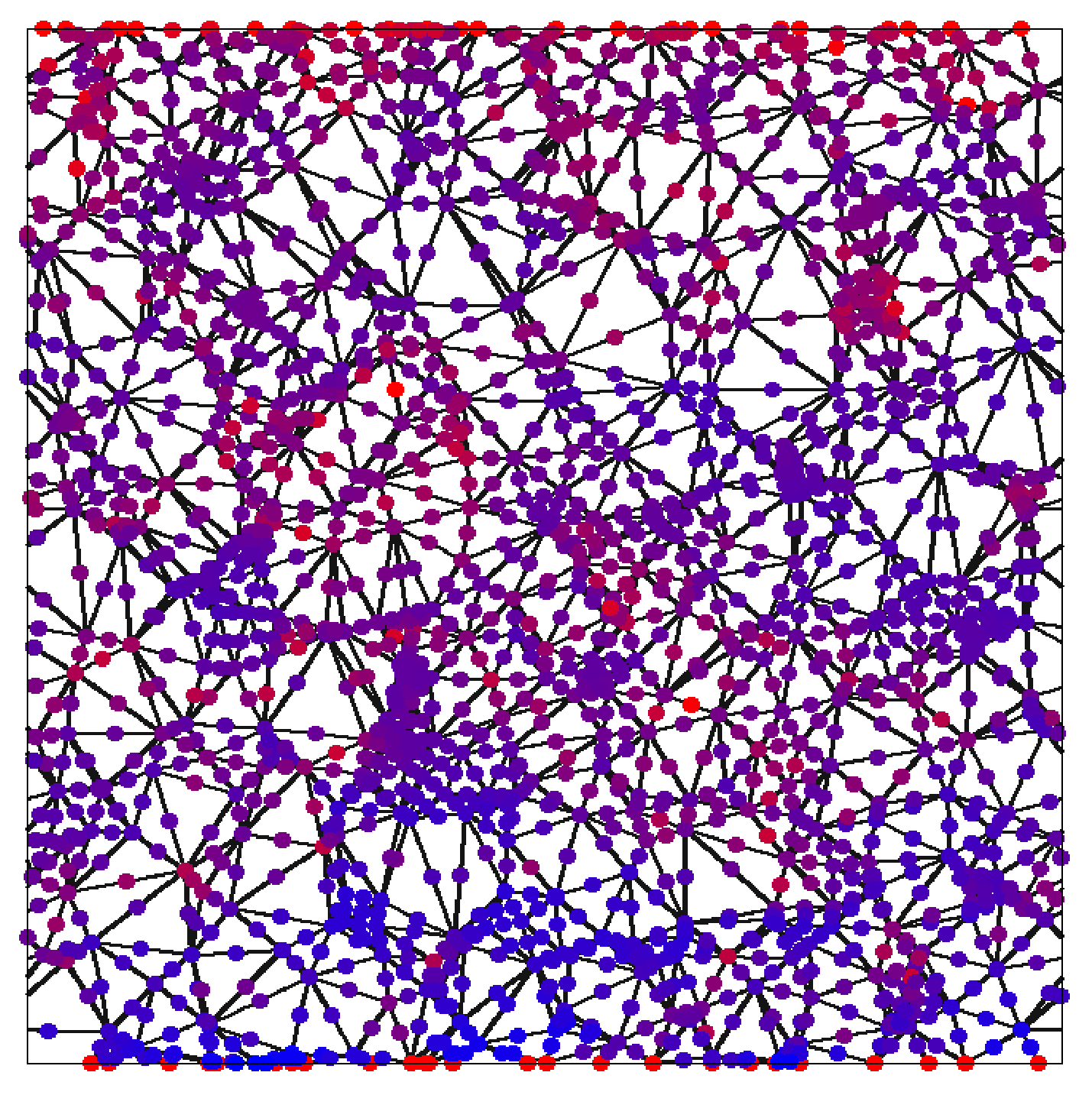}
\label{Delaunay_noeuds}
}
\subfigure[]{
\includegraphics[height=\hei\textwidth]{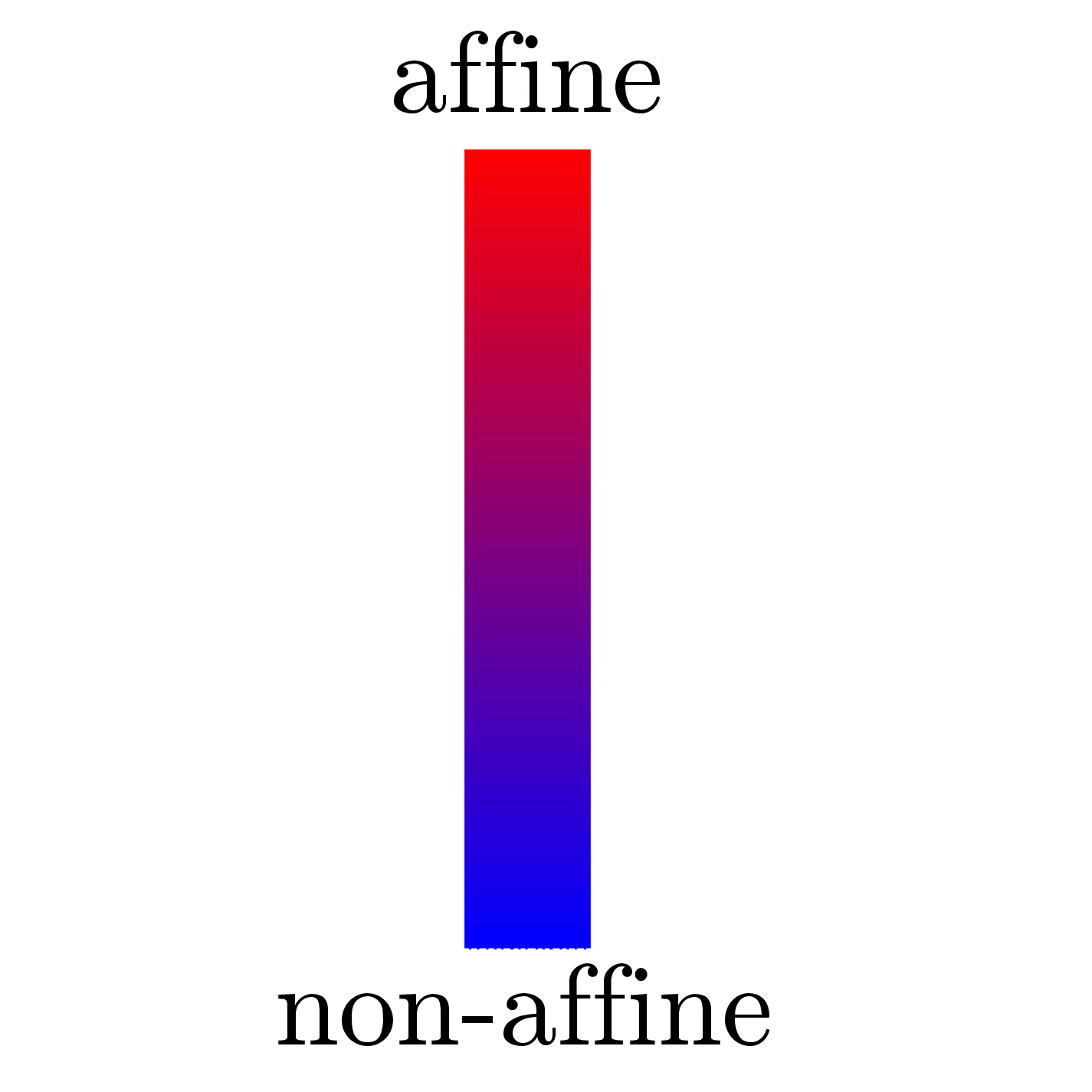}
\label{couleurs_affine}
}
\end{center}
\caption{Affinity measure $m_i = \Vert\mathbf{\bar{u}}_i-\mathbf{u}_i)\Vert/(\gamma L) $ of the displacement field for the same networks as in Figure \ref{repartition_energie} subject to horizontal shear strain. The (logarithmic) calibration bar (f) shows the coloration scheme used for the nodes (junctions and midpoints): they are coloured in red if their displacements match the affine displacement field, and blue in the opposite case.}
\label{affinity}
\end{figure}

Figure \ref{affinity} shows the affinity of the displacement field for the five networks under shear strain. Unsurprisingly, all networks with connectivity $<4$ deform through the bending modes and thus present non-affine displacement fields. On the other hand, the measure of the affinity reveals major differences between networks with connectivity $\geq4$.
Comparison of triangular and Delaunay lattices specially is intriguing: despite the fact that these two networks share similar topological features (they have same mean connectivity and both are fully triangulated), the triangular network has a pure affine strain, while the Delaunay lattice presents a non affine strain. 

Evaluation of the elastic moduli also shows differences between these two networks: we found $\mu/(E_0\phi)\simeq 0.091$ and $M/(E_0\phi)\simeq 0.277$ for Delaunay lattice, while $\mu/(E_0\phi)\simeq 0.125$ and $M/(E_0\phi)\simeq 0.374$ for the triangular lattice (where $\phi$ denotes the network density). Therefore, while both networks are stretch-dominated, the triangular lattice is significantly ($\simeq 27\%$) stiffer than the Delaunay lattice. We checked that the numerical values we obtained were independent of $\kappa_n$,  consistent with our choice for the range of parameter values ($\kappa_n \ll \kappa_s l^2$).
Quite unexpectedly, the kagome and triangular lattices share similar mechanical properties, although these two lattices are structurally very different (kagome has connectivity $z=4$ and is only partially triangulated): both present an affine displacement field, and the values obtained for their elastic moduli are very close (we obtained for kagome lattice: $\mu/(E_0\phi)\simeq 0.125$ and $M/(E_0\phi)\simeq 0.376$). 

These results show that no clear connexion can be established between the rigidity of a network and the (mean) connectivity of its nodes, aside from Maxwell criterion. On the other hand, affinity of the displacement field and network stiffness seems correlated: among the five structures studied, the two networks that deform in an affine way (triangular and kagome lattices) are those with highest elastic moduli. The Delaunay network presents a \textit{piecewise} affine displacement field \cite{note42}, and has significantly lower elastic moduli. Finally the two soft, bending-dominated, networks (hexagonal and Voronoi) have  highly non-affine displacement fields. 
In the next section we provide a theoretical framework to rationalize the relationships between node connectivity, network stiffness, and affinity of the displacement field.

\section{Stiffest networks: existence, structure, elastic moduli \label{Theory}}
%
In the following we show that there exists a class of networks which are stiffer than any other with same symmetry and density (and beam Young's modulus). We restrict our analysis to isotropic structures, although our theoretical framework can be transposed to non-isotropic structures as well. We first establish upper bounds on the elastic moduli of an isotropic network of beams.
These bounds coincide with the numerical values obtained for the triangular and kagome lattices. We then derive the structural properties (actually, a set of necessary and sufficient conditions) of optimal networks, \textit{i.e.} networks that have maximal elastic moduli for a given value of $E_0$ and $\phi$. These conditions involve both geometry and topology, which explain why connectivity alone is not enough to predict the macroscopic behaviour of an elastic network. 
\subsection{Isotropy conditions and bounds on the elastic moduli of isotropic networks}

\begin{figure}[h]
\begin{centering}
\includegraphics[width=0.6\textwidth]{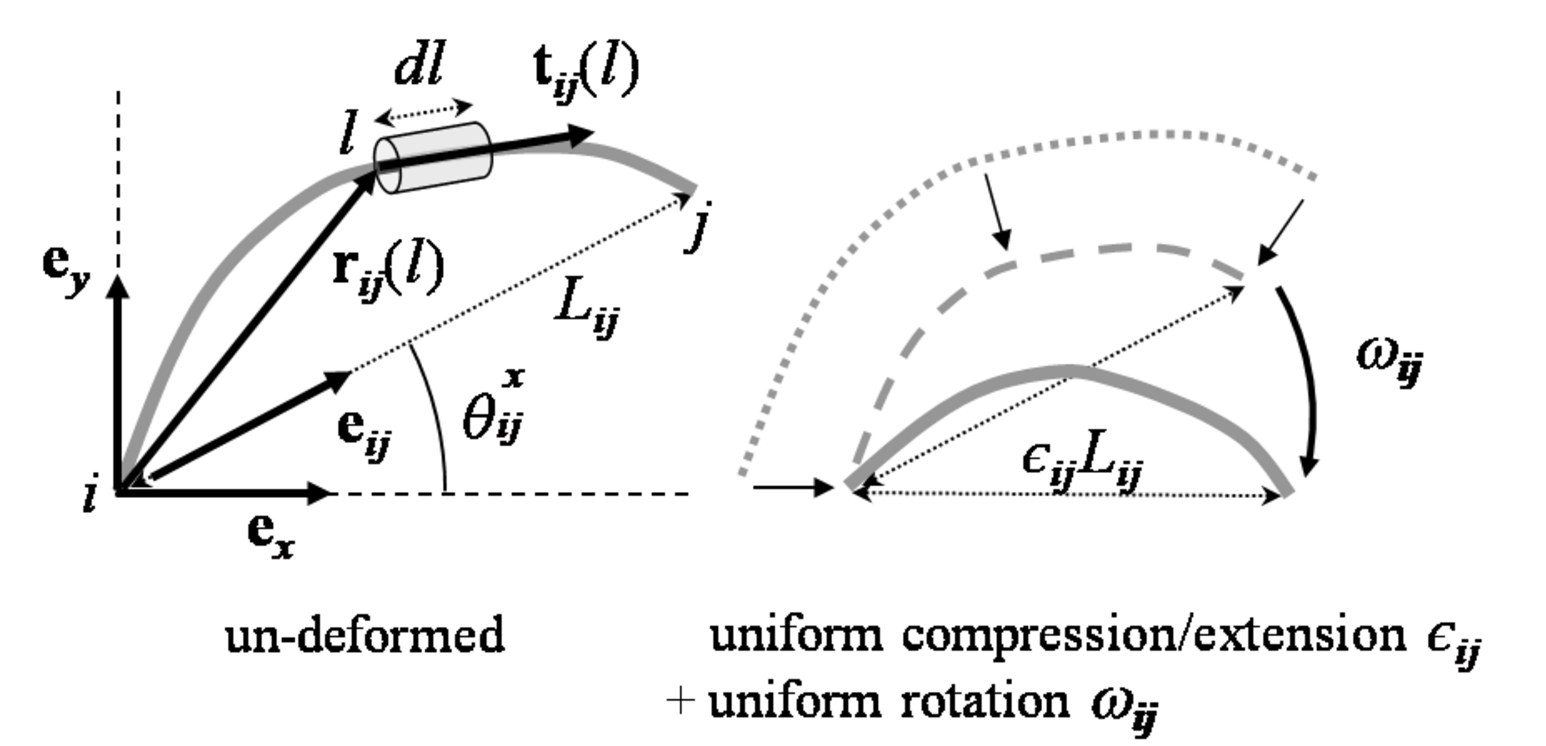}\caption{Geometry of a typical beam
$(i,j)$. In the initial configuration, the beam can have a non-zero natural curvature
and an inhomogeneous cross-section. Under the trial displacement field
(\ref{eq:central_line}), the beam is subjected to a translation $\mathbf{u}_{i}%
$, a uniform compression/extension $\epsilon_{ij}$ and a uniform rotation
$\boldsymbol{\omega}_{ij}$.}%
\end{centering}
\label{notations}%
\end{figure}

Assuming that on the macroscopic scale, the network is a continuous, homogeneous and isotropic medium,
the strain energy per unit volume is (for small strains \cite{Landau}):
\begin{equation}
\varepsilon = \frac{\lambda}{2} (\sum_{\alpha} \bar{u}_{\alpha \alpha})^{2} + \mu \sum_{\alpha,\beta} \bar{u}_{\alpha \beta}^{2},
\label{eq:energy}%
\end{equation}
where $\lambda$ is the Lam\'{e}'s first parameter of the network, $\mu$ the shear modulus (or
Lam\'{e}'s second parameter), $\mathbf{\bar{u}}$ the \textit{macroscopic} displacement field (displacement averaged over a domain large compared to any characteristic homogeneities in the network), and $\bar{u}_{\alpha \beta}=\frac{1}{2} \left(  \frac{\partial
\bar{u}_{\alpha}}{\partial x_{\beta}}+\frac{\partial \bar{u}_{\beta}}{\partial x_{\alpha}}\right)  $ are the
components of the strain tensor ($\bar{u}_{\alpha}$ are the components of  $\mathbf{\bar{u}}\left(  \mathbf{r}\right)  $).
Under uniform loading conditions, $\mathbf{\bar{u}}$ varies linearly with the position $\mathbf{r}$:
$\mathbf{\bar{u}}(\mathbf{r})=\mathbf{A}\centerdot\mathbf{r}$, where $\mathbf{A}$ is a matrix that characterizes the strain. For instance, $A_{\alpha\beta}=\gamma\delta_{\alpha x}\delta_{\beta y}$ for a uniform shear strain $\gamma$ in the $xy$ plan, and $A_{\alpha\beta}=\gamma\delta_{\alpha\beta}$ for a uniform radial strain $\gamma$. 
According to Eq. (\ref{eq:energy}), the density of strain energy corresponding to such homogeneous strain is:
\begin{equation}
\varepsilon=\frac{\lambda}{2}(\sum_{\alpha}a_{\alpha \alpha})^{2}+\mu \sum_{\alpha, \beta}a_{\alpha \beta}^{2}
\label{eq:affineenergy}
\end{equation}
with $a_{\alpha\beta}=\left( A_{\alpha\beta}+A_{\beta\alpha} \right)/2 $.

On the microscopic scale, we suppose that the network is made of interconnected Euler-Bernoulli beams that can have \textit{inhomogeneous natural curvatures and cross-sections}, thus extending our previous analysis on stiff networks \cite{Gurtner-Durand1}. It is worth mentioning that for 3d structures, the Hamiltonian has more terms than in equation (\ref{Hamiltonian}), because of the existence of two bending modes in different planes and the presence of twisting. But since their expressions are not going to appear explicitly in the following, we do not need their precise forms.

The equations of mechanical equilibrium derived from the theory of linear elasticity can be equivalently expressed in terms of a variational principle, sometimes known as the \emph{Principle of
Minimum Potential Energy} (PMPE): consider a body with volume $V$ and
prescribed displacements on its boundaries; the PMPE states that, \emph{among all
kinematically admissible displacement fields (i.e. all continuous displacement 
fields satisfying the displacement constraints on the boundary), the actual
displacement (i.e. the one satisfying the equations of mechanical equilibrium)
is the one that makes the energy functional $\mathcal{E}=\int_{V}$ $\varepsilon
dV$ an absolute minimum}, where $\varepsilon$ is defined through Eq. (\ref{eq:energy}).
This principle is commonly used to find approximate solutions to boundary valued problems (\textit{Rayleigh-Ritz} method). Here we use it to derive rigorous upper bounds on the elastic moduli by choosing an ad-hoc trial displacement field, and then look for the network architecture for which the actual displacement field matches the trial displacement field.

Let us note $\mathbf{r}_{ij}(l)$ and $s_{ij}(l)$
the position vector and cross-sectional area along the beam $(i,j)$, respectively, where $l$ refers to the arc-length starting at node $i$ (see Fig. \ref{notations}). We also note $l_{ij}$ the length of beam $(i,j)$, $L_{ij}$ the distance between nodes $i$ and $j$, and $\mathbf{e}_{ij}$ the unit vector along this straight line.
For any network structure, one can always define a continuous displacement field such that every infinitesimal piece of the beam $(i,j)$ undergoes the same rotation $\boldsymbol{\omega}_{ij}$ and elongation $\epsilon_{ij}$.
This trial displacement field is (for small strains): 
\begin{equation}
\mathbf{u}_{ij}(l)=\mathbf{u}_{i}+\epsilon_{ij}\mathbf{r}_{ij}(l)+\boldsymbol{\omega}_{ij}\times\mathbf{r}_{ij}(l) \;\;\; \forall \; l  \in \left[ 0,l_{ij} \right],
\label{eq:central_line}%
\end{equation}
with $\epsilon_{ij}=\mathbf{e}_{ij}\cdot\left(\mathbf{u}_{j}-\mathbf{u}_{i}\right) / L_{ij}$
and
$\boldsymbol{\omega}_{ij}=\mathbf{e}_{ij}\times\left(  \mathbf{u}_{j}-\mathbf{u}_{i}\right)  /L_{ij}$.
Thus, every beam $(i,j)$ is subject to a translation
$\mathbf{u}_{i}$, a homothety with ratio $\epsilon_{ij}$ and a global rotation
$\boldsymbol{\omega}_{ij}$.
To complete our definition of the
trial displacement field, we impose that each node follows the macroscopic affine displacement: $\mathbf{u}_{i}=\mathbf{A}\centerdot\mathbf{r}_{i}$, where $\mathbf{r}_{i}$ denotes the position vector of node $i$. Clearly, the trial displacement field defined this way is kinematically admissible. 
Since each elementary piece of beam undergoes the same rotation $\boldsymbol{\omega}_{ij}$, 
there is no bending or twisting deformations, and the only
contribution to the strain energy of a piece of beam with infinitesimal length
$\mathrm{d}l$ is the stretching term $\left(  E_{0}/2\right)  s_{ij}(l)\epsilon_{ij}^{2}\mathrm{d}l$. Moreover, for low-density
structures ($s_{ij}^{1/2} \ll l_{ij}$), the node contribution to the strain energy is negligible compared to the typical stretching energy of a beam \cite{note1}.
In this limit, the total energy associated with the trial
displacement field defined above reduces to $\mathcal{E}_{trial}=E_{0}/2
\displaystyle\sum\limits_{(i,j)}v_{ij}\epsilon_{ij}^{2}$, where the summation is
over all the beams that constitute the network, and $v_{ij}=\int_0^{l_{ij}}s_{ij}(l)\mathrm{d}l$ is the volume of
beam $\left(  i,j\right) $.
Introducing the density $\phi$ as the total beam volume per unit area (2d) or unit volume (3d) of the network, the volumetric density of strain energy $\varepsilon_{trial}=\mathcal{E}_{trial}/V$ associated with the trial displacement field reads:
\begin{equation}
\varepsilon_{trial}=\frac{E_{0}}{2} \phi \left\langle \epsilon_{ij}^{2}\right\rangle,
\label{average}
\end{equation}
where the brackets denote the average defined, for any quantity $q_{ij}$, as:
$\left\langle q\right\rangle =\sum_{(i,j)}v_{ij}q_{ij}/\sum_{(i,j)}v_{ij}$.

The elongation $\epsilon_{ij}$ can be rewritten in terms of the elements of $\mathbf{A}$: 

\begin{equation}
\epsilon_{ij}=\mathbf{e}_{ij}\cdot \mathbf{A} \cdot\mathbf{e}_{ij}=\sum_{\alpha,\beta} A_{\alpha\beta} e_{ij}^{\alpha} e_{ij}^{\beta},
\label{eq:extension}
\end{equation}
where $e_{ij}^{\alpha}=$ $\mathbf{e}_{\alpha}\cdot\mathbf{e}_{ij}$ is the
cosine of the angle between the beam $(i,j)$ and the $\alpha$ axis ($\alpha\in\{x,y\}$ for 2d
materials, and $\alpha\in\{x,y,z\}$ for 3d materials). Similarly, the components of the rotation vector $\boldsymbol{\omega}_{ij}=\mathbf{e}_{ij} \times (\mathbf{A} \cdot\mathbf{e}_{ij})$ can be expressed in terms of the elements of $\mathbf{A}$. 

The expression of $\varepsilon_{trial}$ involves
averaged quantities which can be evaluated using symmetry arguments; for
isotropic structures, the strain energy must be identical for any
orientation of the applied strain. Thus, the bound
expressions must remain invariant under rotation around the perpendicular
axes or under permutation of
the axis labels.
After simple but lengthy calculations that we detail in \ref{Appendix}, these invariance properties lead to a
set of relations on the \textit{global} structural properties of optimal networks, which can be summarized as:%

\begin{equation}%
\begin{cases}
\left\langle {e_{ij}^{\alpha}} ^{2}\right\rangle =\frac{1}{d} \qquad
\left\langle {e_{ij}^{\alpha}} ^{4}\right\rangle =\frac{3}{d\left(d+2\right)} & \\
& \\
\left\langle {e_{ij}^{\alpha}} ^{2} e_{ij}^{\beta} e_{ij}^{\gamma} \right\rangle =0 \quad\left(  \beta\neq\gamma\right)  &
\end{cases}
\label{eq:cond_tot}%
\end{equation}
with $\alpha$, $\beta$, $\gamma\in\left\{  x,y\right\}  $ (resp. $\left\{
x,y,z\right\}  $), and $d=2$ (resp. $d=3$) for 2d (resp. 3d) structures. These relations will be referred to as \textit{isotropy conditions} and their consequences will be analysed in the next section. Equations (\ref{eq:cond_tot}) imply in particular that
$\left\langle (e_{ij}^{\beta} e_{ij}^{\gamma})^2\right\rangle
=1/\left(  d\left(  d+2\right)  \right)  $ and $\left\langle e_{ij}^{\beta} e_{ij}^{\gamma}\right\rangle =0$ (for $\beta\neq\gamma$).
Using these relations together with Eqs (\ref{average}) and (\ref{eq:extension}), the density of strain energy simplifies to (see \ref{Appendix}):
\begin{equation}
\varepsilon_{trial}=\frac{E_{0}\phi}{d\left(d+2\right)}\left(\frac{1}{2}(\sum_{\alpha}a_{\alpha \alpha})^{2}+\sum_{\alpha, \beta}a_{\alpha \beta}^{2}\right).
\label{eq:trialenergy}
\end{equation}
According to the PMPE, comparison of Eqs. (\ref{eq:affineenergy}) and (\ref{eq:trialenergy}) yields:%
\begin{equation}
\frac{\lambda}{2}(\sum_{\alpha}a_{\alpha \alpha})^{2}+\mu \sum_{\alpha, \beta}a_{\alpha \beta}^{2}\leq
\frac{E_{0}\phi}{d\left(d+2\right)}\left(\frac{1}{2}(\sum_{\alpha}a_{\alpha \alpha})^{2}+\sum_{\alpha, \beta}a_{\alpha \beta}^{2}\right).
\label{eq:bound}
\end{equation}
Bounds on the elastic moduli can then be deduced from this inequality. Consider first the affine displacement field $A_{\alpha\beta}=\gamma\delta_{\alpha x}\delta_{\beta y}$, corresponding to a shear strain in the $xy$-plane. Inequality (\ref{eq:bound}) reduces to: 
\begin{equation}
\mu \leq \frac{E_{0}\phi}{d\left(d+2\right)}=\mu^\star.
\label{eq:bound_mu}
\end{equation}
Consider then the affine displacement field $A_{\alpha\beta}=\gamma\delta_{\alpha\beta}$, corresponding to a uniform compression. Inequality (\ref{eq:bound}) yields:
\begin{equation}
\kappa \leq \frac{E_{0}\phi}{d^2}=\kappa^\star,
\label{eq:bound_kappa}
\end{equation}
where $\kappa=\lambda+2\mu/d$ denotes the bulk modulus of a $d-$dimensional body.
Any other elastic modulus of an isotropic body is related to $\kappa$ and $\mu$. In particular, Young's modulus reads $E=2 d^2 \kappa \mu/(2\mu+d(d-1)\kappa)$. Since $E$ is an increasing function of $\kappa$ and $\mu$, it comes that:
\begin{equation}
E \leq \frac{2E_{0}\phi}{2+\left(d-1\right)\left(d+2\right)}=E^\star.
\label{eq:bound_E}
\end{equation}

Values of $\mu$, $E$, $M$, and $\kappa$ for 2d and 3d stiffest networks are reported
in table \ref{tableau}.
We also report the values of the Lam\'{e}'s parameter $\lambda$ and Poisson's ratio $\nu$ for this class of networks. However, it must be pointed out that they generally do not correspond to the highest possible values of $\lambda$ and $\nu$. Indeed these two quantities are not elastic moduli, as they are not a direct measure of some strain-stress relationship.
It can be noticed that the Poisson's ratio of such structures is below those of incompressible bodies ($1$ for 2d bodies, $1/2$ for 3d bodies), and is independent of the elastic properties of the beam material ($E_0$).

Values of the elastic moduli reported in Table \ref{tableau} coincide with the Hashin-Shtrikman (HS) bounds for low-density 2d structures \cite{Hashin-Shtrikman2d, Torquato}, but are strictly lower (i.e. tighter) for 3d structures \cite{Hashin-Shtrikman} (see Table \ref{tableauHS}). However, it must be pointed out that this improvement of the HS upper-bounds is restricted to materials whose elastic phase is organized into slender objects (beams), while the HS bounds apply to any diphasic structures, including for instance those where the continuous phase in 3d structures is assembled into 2d sheets.

\begin{table}[ptb]
\centering
\newcolumntype{Y}{>{\centering\arraybackslash}X}
\begin{tabularx}{\linewidth}[c]{|Y|Y|Y|Y|Y||Y|Y|}\hline
&  $\mu^\star/E_{0}$ & $E^\star/E_{0}$   & $\kappa^\star/E_{0}$ & $M^\star/E_{0}$ & $\lambda^\star/E_{0}$ & $\nu^\star$\\\hline
\textbf{2d} & $\phi/8$ & $\phi/3$ & $\phi/4$ & $3\phi/8$ & $\phi/8$ & $1/3$\\\hline
\textbf{3d} & $\phi/15$ & $\phi/6$  & $\phi/9$ & $\phi/5$ & $\phi/15$ & $1/4$\\\hline
\end{tabularx}
\caption{Elastic moduli of stiffest isotropic structures normalized by the
Young's modulus of the beam material $E_{0}$: ($\mu$), Young's modulus ($E$),
bulk modulus ($\kappa$), longitudinal modulus ($M$). We also report the Lam\'{e}'s first parameter ($\lambda$) and Poisson's ratio ($\nu$) of such optimal structures; note however that these two quantities are not elastic moduli, so $\lambda^\star$ and $\nu^\star$ do not necessarily correspond to their highest possible values.}%
\label{tableau}%
\end{table}

\begin{table}[ptb]
\centering
\newcolumntype{Y}{>{\centering\arraybackslash}X}
\begin{tabularx}{\linewidth}[c]{|c|c|c|c|c|}\hline
&  $\mu^{(HS)}/E_{0}$ & $E^{(HS)}/E_{0}$   & $\kappa^{(HS)}/E_{0}$ & $M^{(HS)}/E_{0}$ \\\hline
\textbf{2d} & $\phi/8$ & $\phi/3$ & $\phi/4$ & $3\phi/8$\\\hline
\textbf{3d} & $\dfrac{\phi}{15} \left( 1+ \right.$ & $\dfrac{\phi}{6} \left( 1+3\dfrac{49\kappa_0 +8\mu_0}{69\kappa_0+8\mu_0} \right)$  & $\dfrac{\phi}{9} \left(1+\dfrac{9\kappa_0 }{3\kappa_0+4\mu_0} \right)$ & $\dfrac{\phi}{5} \left( 1+ \right.$ \\
 & $\left. \dfrac{18\kappa_0^2 +8\mu_0^2+21\kappa_0 \mu_0}{3\kappa_0\left(3\kappa_0+4\mu_0\right)} \right)$ & & & $\left. \dfrac{141\kappa_0^2 +4\mu_0^2+116\kappa_0 \mu_0}{9\kappa_0\left(3\kappa_0+4\mu_0\right)} \right)$
\\\hline
\end{tabularx}
\caption{Hashin-Shtrikman upper-bounds on the elastic moduli of low-density isotropic materials with one single elastic phase \cite{Hashin-Shtrikman2d,Torquato,Hashin-Shtrikman}: shear modulus($\mu^{(HS)}$), Young's modulus ($E^{(HS)}$), bulk modulus ($\kappa^{(HS)}$), longitudinal modulus ($M^{(HS)}$). The bounds are normalized by $E_0$. The bounds derived in the present paper and reported in Table \ref{tableau} coincide with the HS bounds for low-density 2d structures, and are strictly lower (i.e. tighter) for 3d structures.}%
\label{tableauHS}%
\end{table}
\subsection{Mechanical conditions}
We now show that inequalities (\ref{eq:bound})-(\ref{eq:bound_E}) become strict equalities
for some specific network geometries, determined by the following set of rules
(along with the isotropy conditions (\ref{eq:cond_tot})):

 \textbf{(a)} All the beams are straight.

 \textbf{(b)} Every beam $(i,j)$ has uniform cross-sectional area: $s_{ij}%
(l)=s_{ij}$.

 \textbf{(c)} At every junction $i$ of the 2d (resp. 3d) network, and for all
$\mathbf{e}_{\alpha}$, $\mathbf{e}_{\beta}$, $\mathbf{e}_{\gamma}\in\left\{
\mathbf{e}_{x},\mathbf{e}_{y}\right\}  $ (resp. $\left\{  \mathbf{e}%
_{x},\mathbf{e}_{y},\mathbf{e}_{z}\right\}  $), the following equality is
satisfied:
$$
\sum\limits_{j\in \mathcal{N}(i)}s_{ij}\e{\alf}\e{\bet}\e{\gam}=0, \label{eq:cond_forces}
$$
where $\mathcal{N}(i)$ denotes the set of neighbouring nodes that are connected to node $i$.
Unlike the isotropy conditions (\ref{eq:cond_tot}), these mechanical conditions are local rules on the geometry and topology of the network.

The demonstration is straightforward: according to the PMPE, the inequalities
(\ref{eq:bound})-(\ref{eq:bound_E})
become strict equalities \emph{if and only if} the respective trial displacement fields coincide with the displacement field that satisfies the equations of mechanical equilibrium. 
Inspection of force and moment balances along each beam and at each junction leads to the three necessary and sufficient conditions stated above: 
with the trial displacement field (\ref{eq:central_line}), each infinitesimal piece of beam is
subjected to an axial deformation only. Therefore, the force and local moment
deriving from the trial displacement fields are $\mathbf{F}_{ij}\left(
l\right)  =-E_{0}s_{ij}\left(  l\right)  \epsilon_{ij}\mathbf{t}_{ij}\left(
l\right)  $ and $\mathbf{M}_{ij}\left(  l\right)  =\mathbf{0}$, respectively, where $\mathbf{t}_{ij}\left(  l\right) = \dd \mb{r}_{ij}/\dd l$ is the local tangent unit vector ($\mathbf{F}_{ij}\left(  l\right)  $ and $\mathbf{M}_{ij}\left(  l\right)  $
are defined as the force and local torque exerted at position $l$ by the
$i$-side to the $j$-side of beam $(i,j)$). The force balance equation
$d\mathbf{F}_{ij}\left(  l\right)  /dl=\mathbf{0}$ along the beam $\left(
i,j\right)  $ yields:
\begin{equation}
\dfrac{\dd s_{ij}}{\dd l}(l)\mb{t}_{ij}(l)+s_{ij}(l)\dfrac{\dd \mb{t}_{ij}}{\dd l}=\mb{0}.
\end{equation}
Since $\mb{t}_{ij}(l)$ and $\dd \mb{t}_{ij}/\dd l$ are orthogonal vectors, it comes that $\dd s_{ij}(l)/\dd l=0$ and $\dd\mb{t}_{ij}/\dd l=\mb{0}$ for all $l \in \left[ 0,l_{ij} \right]$, what immediately leads to condition (a) and (b). The moment balance equation $d\mathbf{M}_{ij}\left(  l\right)  /dl + \mb{t}_{ij}(l) \times \mathbf{F}_{ij}\left(  l\right)=\mathbf{0}$ is automatically satisfied for such geometry.

Mechanical equilibrium must be satisfied at every junction $i$ as well. When condition (a) and (b) are fulfilled, the force exerted by the straight and uniform beam $\left(  i,j\right)  $ on node $i$ is
$E_{0}s_{ij}\epsilon_{ij}\mathbf{e}_{ij}$. Thus, no torque is exerted on the junction, while the balance of forces yields: $\sum_{j\in \mathcal{N}(i)}s_{ij}\epsilon_{ij}\mathbf{e}_{ij}=\mathbf{0}$. This
relation must hold for any orientation of the strain field. 
Replacing $\epsilon_{ij}$ by its expression (\ref{eq:extension}) and using the same rotational invariance arguments than in (\ref{isotropy_conditions}) eventually yields condition (c).

\section{Analysis of the isotropy and mechanical conditions \label{Analysis}}
\subsection{Isotropy conditions \label{isotropy_conditions}}
The isotropy conditions (\ref{eq:cond_tot}) constitute a set or 4 (resp. 15) conditions for 2d (resp. 3d) structures. 
We first check that these conditions are satisfied with a continuous and uniform angular distribution of beams (such a distribution is rather unrealistic, but it must have isotropic properties by construction). For that purpose, we reformulate the average $\< \centerdot \>$ defined in Eq. (\ref{average}) for the case of a continuous distribution: let us note $f\left(v,\mathbf{n}\right)$ the probability density function for a beam to have a volume $v$ and an orientation along the unit vector $\mathbf{n}$.
Then, the average $\< \centerdot \>$ of any quantity $X(v,\mb{n})$ is:
\begin{equation}
\< X\> = \sum_{(i,j)} \frac{v_{ij}}{V} X_{ij} = \frac{N}{V} \int_V \int_\Omega v f(v,\mb{n}) X(v,\mb{n})\dd v\dd\Omega,
\label{continuous_average} 
\end{equation}
where $N$ is the total number of beams within the network, $V$ their total volume, and $\dd\Omega$ denotes the elementary angle (2d) or solid angle (3d) of orientation $\mb{n}$. If the distribution $f$ is isotropic (\textit{i.e.:} $f$ is independent of $\mb{n}$), and the quantity $X$ depends only on the orientation $\mb{n}$ (\textit{e.g.:} $X_{ij}=\cos^2\theta_{ij}$), Eq. (\ref{continuous_average}) simplifies to
\begin{equation}
\< X\> = c \int_\Omega  X(\mb{n}) \dd \Omega.
\label{continuous_distrib}
\end{equation}
The constant $c$ is obtained by normalization: $c=\overline{v}/\pi$ (resp. $c=\overline{v}/(2 \pi)$) for 2d (resp. 3d) networks, where $\overline{v}$ is the mean beam volume in the network.
Using Eq. (\ref{continuous_distrib}), it is straightforward to see that a uniform distribution of beams satisfies the isotropy conditions (\ref{eq:cond_tot}). 
More realistic, discrete distributions of beam orientation can also satisfy the isotropy conditions.
In two dimensions, conditions (\ref{eq:cond_tot}) can be rewritten as:
\begin{align}
\left\langle  \cos 2\theta_{ij} \right\rangle =0, & &
\left\langle  \sin 2\theta_{ij} \right\rangle =0, & &
\left\langle  \cos 4\theta_{ij} \right\rangle =0, & &
\left\langle  \sin 4\theta_{ij} \right\rangle =0,
\label{eq:isotropy_2d}
\end{align}
where $\theta_{ij}$ is the angle between the beam $(i,j)$ and the $x$-axis. We can draw some general properties for networks with uniform angular distribution of beams, \textit{i.e.} such that the total volume of beams with same (discrete) orientation $\theta_{m}$ is independent of $\theta_{m}$ (special cases of such networks are the periodic frameworks with identical beams and with junctions which are all similarly situated \cite{Deshpande}).
Using the complex variables $y_{m}=\cos \theta_{m}+\imath \sin \theta_{m}$, the equations (\ref{eq:isotropy_2d}) reduce to:
\begin{align}
\sum_{\lbrace m \rbrace} y_{m}^2=0, & &
\sum_{\lbrace m \rbrace} y_{m}^4=0,
\label{eq:complex_isotropy_2d}
\end{align}
where the summation is over all the distinct beam orientations. 
From this set of equations, it is clear that a distribution with two distinct beam orientations cannot have isotropic properties: square or rectangular lattices, for examples, are known to have anisotropic elastic properties \cite{Landau}. On the other hand, one can build isotropic networks with three sets of parallel beams. A simple analysis of Eqs (\ref{eq:complex_isotropy_2d}) shows that the three sets of beams must be tilted from each other with equal angles (modulo $\pi$) of $\pi/3$. Hexagonal, triangular, and kagome lattices are examples of such isotropic structures.
The analysis of the mechanical conditions below allows to determine which ones of them have highest stiffness.
\subsection{Mechanical conditions}
The condition (c) produces a set of 4 (resp. 10) equations
per node for 2d (resp. 3d) structures, imposing severe restrictions on the geometry and topology of a junction. 
In this section we inspect the solutions to this set of equations for 2d networks. We suppose that conditions (a) and (b) are satisfied (straight and uniform beams). Moreover, we assume that beams have the same cross-section.
Let $z_{i}$ be the connectivity of node $i$, and $\theta_{ij}$ the angle between the beam $(i,j)$ and the $x$ axis. The vectorial condition (c) can be rewritten as $4$ trigonometric relations:
\begin{align}
 \sum_{j=1}^{z_{i}} \cos\theta_{ij}=0, & &
 \sum_{j=1}^{z_{i}} \sin\theta_{ij}=0, & &
 \sum_{j=1}^{z_{i}} \cos3\theta_{ij}=0, & &
 \sum_{j=1}^{z_{i}} \sin3\theta_{ij}=0.
 \label{eq:cos_et_sin}
\end{align}

\begin{figure}[htbp]
\begin{center}
\includegraphics[width=0.8\textwidth]{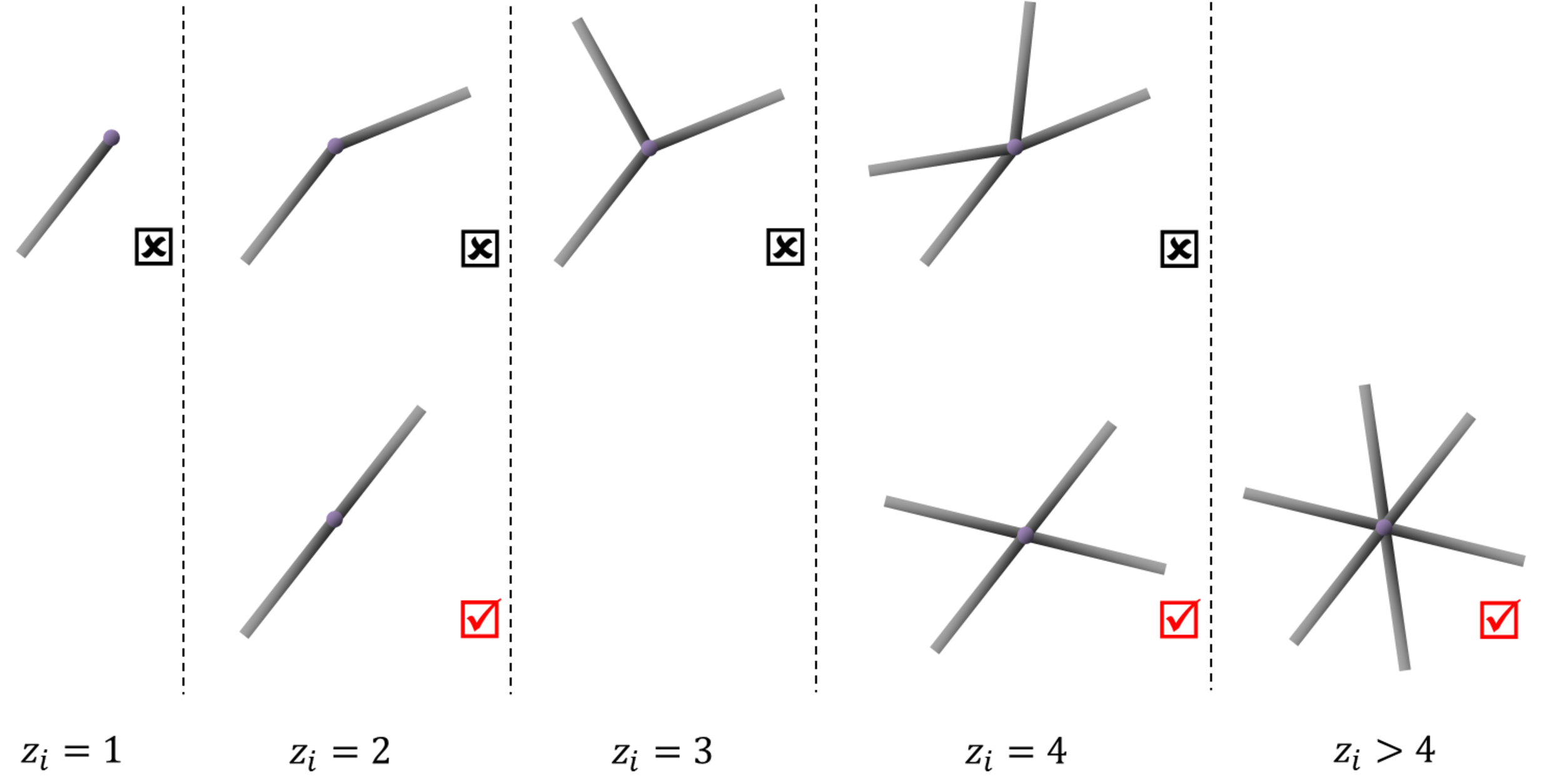}
\end{center}

\caption{Possible geometries for a junction with connectivity $z_i$ in an optimal (and isotropic) network, with beams having the same-cross-section. The allowed geometries are checked in red. Configurations where beams are parallel in pairs are always solutions to Eqs (\ref{eq:isotropy_2d}) and (\ref{eq:cos_et_sin}). These are the only solutions when $z_i\leq4$.}
\label{fig:junctions}
\end{figure}

Clearly, there is no solution to this set of equations for a monovalent node ($z_{i}$=1): all the beams of an optimal network are connected at their both ends and can contribute to the storage of elastic energy.
For a divalent node ($z_{i}$=2), the only solutions are the trivial configurations $\theta_{i2}=\theta_{i1}+\pi$. To inspect the possible configurations of nodes with higher connectivity, we rewrite Eqs (\ref{eq:cos_et_sin}) as: 

\begin{align}
\sum_{j=1}^{z_{i}} y_{ij}=0, & &
\sum_{j=1}^{z_{i}} y_{ij}^3=0,
\label{eq:complex-cos_et_sin}
\end{align}
where we introduced the complex variables $y_{ij}=\cos\theta_{ij}+\imath \sin\theta_{ij}$.
One can now show that there is no solution to Eqs (\ref{eq:complex-cos_et_sin}) for a trivalent node ($z_{i}$=3): eliminating 
$y_{i3}$ from these equations 
yields $y_{i1} y_{i2} (y_{i1}+y_{i2})=0$; either $y_{i1}$, $y_{i2}$ or $y_{i3}$ is zero, what is not compatible with the normalization constraint  $\vert y_{ij}\vert =1$.
For quadrivalent nodes ($z_{i}$=4), eliminating $y_{i4}$ from Eqs (\ref{eq:complex-cos_et_sin}) yields:
$(y_{i1}+y_{i2})(y_{i2}+y_{i3})(y_{i1}+y_{i3})=0$. Thus, the only possible configurations for such nodes are $\theta_{i3}=\theta_{i1}+\pi$, $\theta_{i4}=\theta_{i2}+\pi$, and all other subscript permutations, \textit{i.e.} beams must be collinear in pairs.
In agreement with Maxwell's criterion \cite{Deshpande, Dunlop, Thorpe3, Heussinger2}, it comes that 2d stiff networks must have a node connectivity $\geq 4$. But in addition, our analysis specifies what must be the geometry of the junctions (see Fig. \ref{fig:junctions}). In the special case where nodes are all similarly situated, it can then be shown that the minimal node valency is $6$ \cite{Deshpande}. It is possible, in principle, to find the solutions to Eqs (\ref{eq:cos_et_sin}) for nodes with higher valencies. The solutions are certainly more tedious to find. However, it can be mentioned that for junctions with an even number of adjoining beams, the geometry with beams parallel in pairs is always a solution to Eqs (\ref{eq:cos_et_sin}).

As a consequence of our analysis, it comes that 2d optimal networks must contain triangular cells (but the converse is generally not true, \textit{e.g.} see Delaunay network in Fig. \ref{Delaunay_noeuds}); the number of cells $C$, the number of sides (beams) $N$, and the number of junctions $J$ of a 2d network are related by the Euler's formula: $C-N+J=1$ \cite{Deshpande}. Therefore, the mean coordination number $\bar{z} = 2N/J$ and the mean number of sides per cell $\bar{n} = 2N/C$ satisfy: $1/\bar{n} + 1/\bar{z} = 1/2$.
Thus, in optimal networks, $\bar{z} \geq 4$ and $\bar{n} \leq 4$. Clearly, optimal networks contain triangles when $\bar{z} > 4$. The marginal case $\bar{z}=4$ ($\bar{n}=4$) must be inspected separately: either each cell has exactly 4 sides; in order to satisfy the mechanical condition, every vertex must be a fourfold junction with beams collinear in pairs, but we have seen that a structure with only two distinct beam orientations does not satisfy the isotropy conditions. Or cells do not have all the same number of sides; then the constraint $\bar{n}=4$ imposes that such a network must contain triangles. 

A striking feature of the stiffest networks is that, under uniform loading, the displacement field is affine down to the microscopic scale and thus coincides with the macroscopic displacement field $\bar{\mathbf{u}}$: since every beam must be straight, one has $\mathbf{r}_{ij}(l)=l\mathbf{e}_{ij}$. Thus, according to Eq. (\ref{eq:central_line}), the displacement field at location $\mathbf{r}=\mathbf{r}_i+\mathbf{r}_{ij}(l)$ simplifies to $\mathbf{u} (\mathbf{r})=\mathbf{A}\centerdot\mathbf{r}_{i}+\mathbf{A}\centerdot l\mathbf{e}_{ij}=\mathbf{A}\centerdot\mathbf{r}=\bar{\mathbf{u}}(\mathbf{r})$.

Finally, it is worth mentioning that isotropic networks with optimal elastic properties also have optimal transport properties. Indeed, the isotropy conditions (\ref{eq:cond_tot}) encompass the isotropy conditions $\langle e_{ij}^{\alpha} e_{ij}^{\beta} \rangle =\delta_{\alpha,\beta}/d$ for transport properties. Furthermore, using the identity $\sum_{\alpha=1}^{d} \left( e_{ij}^{\alpha} \right)^2=1 $, it is easy to show that condition (c) implies $\sum_{j}s_{ij}\mathbf{e}_{ij}=\mathbf{0}$, which is the condition required at every junction of a network (together with conditions (a) and (b)) to have optimal \textit{transport} properties \cite{Durand1,Durand2}. 

\subsection{Comparison with numerical results}
The analysis of the isotropy and mechanical conditions made above sheds light on the numerical results reported in section \ref{Simulations}: the five simulated networks satisfy the isotropy conditions (\ref{eq:isotropy_2d}), but only the kagome and triangular networks satisfy the mechanical conditions. It is noteworthy that the kagome lattice is one of the stiffest networks, in spite of its large number of floppy modes \cite{Hutchinson1,Hutchinson2,Sun}. However, small defects in its structure will dramatically affect its mechanical response \cite{Gurtner-Durand3}.

The mechanical conditions involve both geometry (orientation and cross-sectional areas of the beams) and topology (connectivity) of each junction. That explains why two elastic networks with same connectivity, like triangular and Delaunay lattices, can have very different macroscopic responses.
Moreover, our analysis has revealed the strong correlation between the stiffness of a network and the affinity of its displacement field. This is also what we observe in the simulations: the two networks with highest elastic moduli -- kagome and triangular lattices -- are also the only ones with pure affine deformations.
It must be emphasized that all stretch-dominated networks do not deform in an affine way.
This is illustrated in our simulations with the Delaunay lattice: this structure deforms exclusively through the stretching of its members (see Fig. \ref{Delaunay_poutres}), but does not deform affinely \cite{note42} (see Fig. \ref{Delaunay_noeuds}). Indeed, the condition (c) is not satisfied in a Delaunay lattice.

\section{Examples of optimal structures} \label{Examples}
In sections \ref{Theory} and \ref{Analysis}, we have derived the structural conditions that an isotropic network must fulfilled to have highest elastic moduli for a given density and Young's modulus of beam material. Nevertheless, one might wonder whether such networks do exist, that is, whether one can effectively build networks that satisfy all these conditions simultaneously. The answer is obviously yes in two dimensions, as we already provided two examples of optimal networks (kagome and triangular lattices). In this section we show how our theoretical analysis can be used to identify other optimal networks in both two and three dimensions. 

\begin{figure}[htbp]
\subfigure[]{
\includegraphics[width=0.46\textwidth]{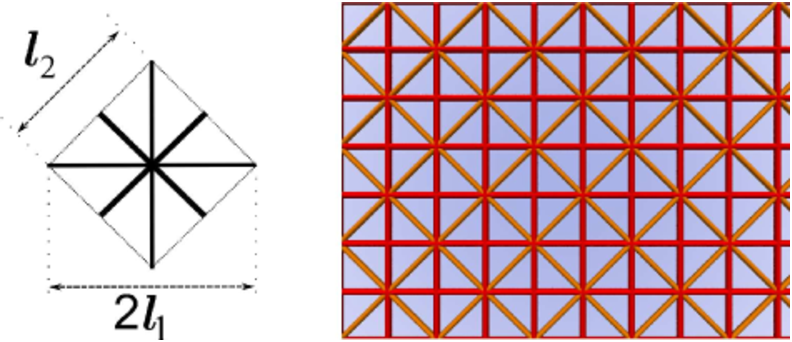}
\label{fig:los}
}
\subfigure[]{
\includegraphics[width=0.48\textwidth]{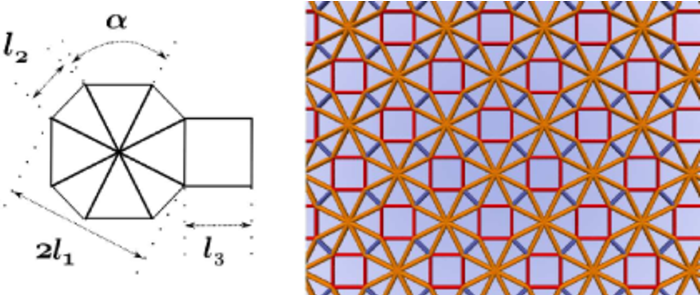}
\label{fig:octo}
}
\subfigure[]{
$\quad \quad$
\includegraphics[width=0.41\textwidth]{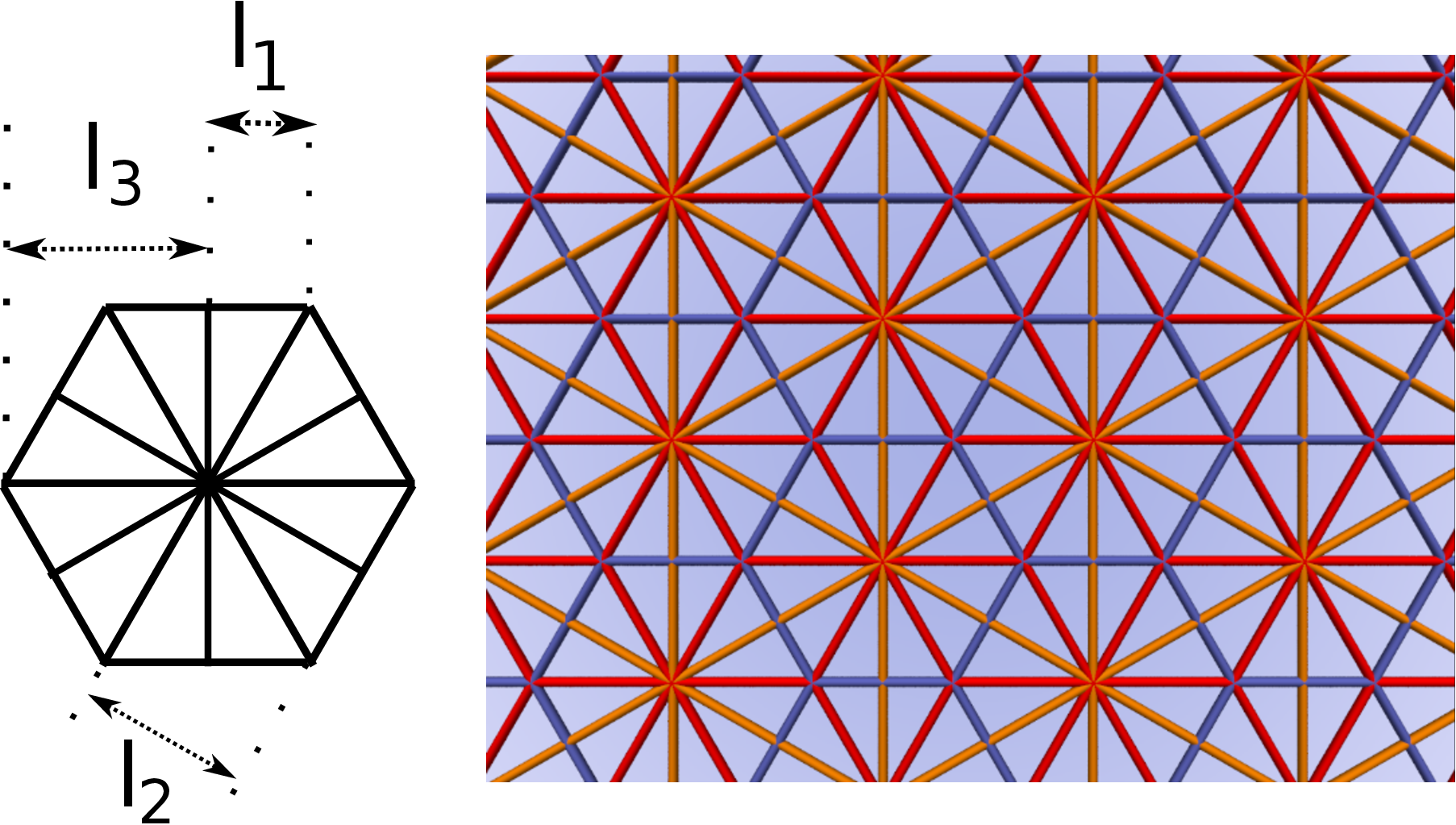}
\label{fig:hexakis}
}
\label{fig:2Dexamples}
\caption{Optimal networks with beams of different sizes (lengths $l_i$ and cross-sections $s_i$); (a): Union Jack lattice, made with two different kinds of beams. The optimality is obtained for $s_2/s_1=\sqrt{2}$; (b): network with three different kinds of beams. The optimality is obtained for $\alf\simeq 53.15^\circ$, $s_2/s_1 \simeq 0.506$ and $s_3/s_1 \simeq 0.805$; (c): network based on the kisrhombille tiling and containing three different kinds of beams. The optimality is reached when $s_3=s_1$, whatever the value of $s_2$.}

\end{figure}

Since isotropy and mechanical conditions are explicit, it is easy to use them to find new optimal networks. Figure \ref{fig:2Dexamples} shows three additional optimal networks with 2d periodic arrangements, here made of non-identical beams. The first one, often referred to as the \textit{Union Jack lattice}, is made with two different kinds of beams. Beams are collinear in pairs at every junction, so the mechanical conditions are satisfied. The ratio of beam cross-sections is then adjusted to fulfil the 2d isotropy conditions (\ref{eq:isotropy_2d}): the only restrictive condition is $\< \cos 4\theta\>=0$ \cite{note4}, yielding $2(l_1 s_1-l_2 s_2) = 0$ (see notations on Fig. \ref{fig:los}). Since, from geometry, $l_2=l_1/\sqrt{2}$, it comes that $s_2/s_1=\sqrt{2}$. 

The second network (Fig. \ref{fig:octo}) is built with three different kinds of beams. Here beams are not collinear in pairs at every junction. The isotropy conditions $\< \sin 2\theta\>=\< \cos 2\theta\>=0$ are satisfied. The angle $\alpha$ and the two cross-section ratios $r_2=s_2/s_1$ and $r_3=s_3/s_1$ (see notations in Fig. \ref{fig:octo}) are then adjusted to satisfy the remaining isotropy and mechanical conditions. We obtain:
\begin{align}
\displaystyle r_3=\frac{r_2}{\sqrt{2}} -\sin\frac{\alf}{2}+\cos\frac{\alf}{2}, & &
\displaystyle r_2=\sqrt{2}\, \frac{1-\sin\alf-\cos\alf-2\cos2\alf}{3\sin \left(\alf/2\right)-\cos\left(\alf/2\right)}, & &
\displaystyle r_2=\sqrt{2} \sin\alf \left(\cos\frac{\alf}{2}-\sin\frac{\alf}{2}\right).
\end{align}
This set of equations can be solved numerically; we obtain: $\alf \simeq 0.927 $ rad $\simeq 53.15^\circ$, $r_2 \simeq 0.506$, $r_3=0.805$. It can be noticed that, like the kagome lattice, this optimal structure is not fully triangulated.

The third network (Fig. \ref{fig:hexakis}), based on the kisrhombille tiling, is made of three different kinds of beams. From geometry, $l_2=\sqrt{3}l_1$ and $l_3=2l_1$. As before, the isotropy conditions are satisfied by construction, and the cross-section ratios $r_2$ and $r_3$ are adjusted to satisfy the mechanical conditions. One obtains that the optimality is reached when $r_3=1$, whatever the value of $r_2$.

We also performed simulations of these networks and checked that the numerical values of their elastic moduli are in very good agreement with the expected values: $\mu/(E_0\phi)\simeq 0.125$, $M/(E_0\phi)\simeq 0.375$ for the network of Fig. \ref{fig:los}, $\mu/(E_0\phi)\simeq 0.124$, $M/(E_0\phi)\simeq 0.374$ for the network of Fig. \ref{fig:octo}, and
$\mu/(E_0\phi)\simeq 0.125$, $M/(E_0\phi)\simeq 0.375$ for the network of Fig. \ref{fig:hexakis}.

\begin{figure}
\begin{center}
\includegraphics[width=0.8\textwidth]{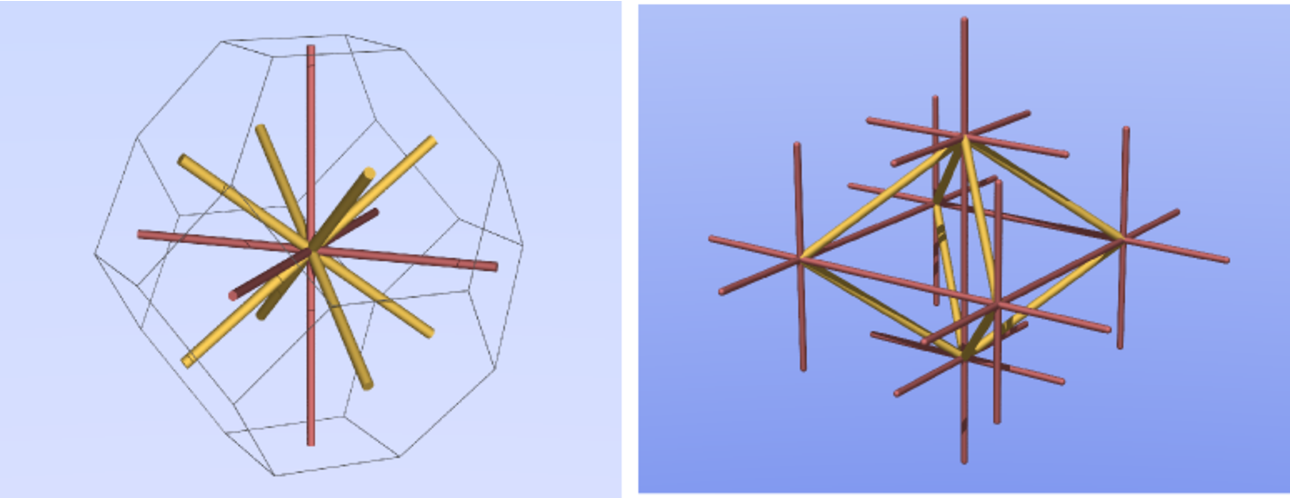} 
\end{center}
\caption{Example of 3d optimal structure with periodic arrangement. Left: a single mesh is inscribed in a Kelvin cell, and thus tiles the space. It has two different sorts of beams, with respective length and cross-section $l_1$, $s_1$ (in red) and $l_2$, $s_2$ (in yellow). The mechanical conditions are satisfied by construction, and the isotropy conditions (\ref{eq:cond_tot}) are fulfilled for $s_{2}/s_{1}=3\sqrt{3}/4$. Right: illustration of the structure obtained from tiling with this unit cell. One can note that it forms two entangled cubic networks (in red), linked to each other by transverse beams (in yellow).}
\label{fig:3dstructure}
\end{figure}

The three-dimensional case is certainly more complex. 
As a matter of fact, we were not able to identify
any periodic and isotropic optimal structure made with one single kind of beam.
However, stiffest structures built with two (or more) types of beams do exist. For instance, the
structure depicted on Fig. \ref{fig:3dstructure} satisfies the mechanical
conditions and can be repeated periodically: the bemas join the centres of the
faces of a Kelvin cell (which is known for tiling the space \cite{Weaire}). The length ratio
of the two kinds of beams is imposed by the Kelvin cell geometry: $l_{2}%
/l_{1}=$ $\sqrt{3}/2$. We then adjust the section ratio for the isotropy
conditions (\ref{eq:cond_tot}) to be satisfied.\ We obtain: $s_{2}/s_{1}%
=3\sqrt{3}/4$.
To our knowledge, this optimal (and isotropic) network with periodic arrangement has never been published before and represents one of the main result of the present study.

\section{Conclusions}
In summary, we showed the existence of a class of isotropic networks having highest possible values of elastic moduli for a given density. We established a convenient set of conditions that allow to identify these optimal networks. The elastic moduli of these networks are also derived, and can be simply expressed in terms of the Young's modulus of the beam material and the
relative density of the structure. Examples of two- and three-dimensional
optimal structures with periodic arrangements are also given. These results may be of interest for structural applications as well as for our
understanding of biological systems. Moreover, they can be used to study the mechanical properties of any (isotropic) structure which can be viewed as a slightly perturbed optimal network \cite{Gurtner-Durand3}.

\appendix
\section{Isotropy conditions} \label{Appendix}
Introducing the expression (\ref{eq:extension}) of the elongation $\epsilon_{ij}$ in Eq. (\ref{average}) yields  
\begin{eqnarray}
\varepsilon_{trial}&=&\frac{E_{0} \phi}{2}\left(\sum_\alpha a_{\alf\alf}^2 \< \e{\alf}^4 \>\right. + \sum_{\alf\neq\bet} \left(a_{\alf\alf}a_{\bet\bet} + 2 a_{\alf\bet}^2\right) \<\e{\alf}^2\e{\bet}^2\> \nonumber \\
				   & &+ 4\sum_{\alf\neq\bet} a_{\alf\alf}a_{\alf\bet} \<\e{\alf}^3\e{\bet}\> + \left. 2 \sum_{\alf\neq\bet,\alf\neq\gam,\bet\neq\gam}  \left(
a_{\alf\alf}a_{\bet\gam}+2 a_{\alf\bet}a_{\alf\gam}\right) \< \e{\alf}^2\e{\bet}\e{\gam}\> \right), \label{eq:fourth}
\end{eqnarray}
where $a_{\alpha \beta}=(A_{\alpha\beta}+A_{\beta\alpha})/2$. For an isotropic network, the expression (\ref{eq:fourth}) must be invariant by permutation or inversion of axes, for any applied strain. These invariance properties lead to restrictions on the structure of the network, and simplify the expression of $\varepsilon_{trial}$. 

First we note that the last sum in Eq. (\ref{eq:fourth}) vanishes for 2d networks, as $\alf$, $\bet$ and $\gamma$ cannot be all distinct. We show that this sum also vanishes for 3d isotropic networks, by analysing the strain field defined as: $a_{xx}=a_0$, $a_{yz}=a_{zy}=b_0$, and the other components $a_{ij}=0$. Thus Eq. (\ref{eq:fourth}) reduces to $\varepsilon_{trial} =  (E_0 \phi/2) \left( a_0^2 \<{\e{x}}^4\> +4 b_0^2 \<{\e{y}}^2{\e{z}}^2\> + 4a_0 b_0 \<{\e{x}}^2\e{y}\e{z}\>\right)$. Inverting the  $y$ or $z$ axes changes the sign of the last term only. But as $\varepsilon_{trial}$ must remain unchanged under such an operation, one necessarily has $\< \e{x}^2\e{y}\e{z}\>=0$, and more generally, by permutation of the axes: $\< \e{\alf}^2\e{\bet}\e{\gamma}\>=0$ with $\alf$, $\bet$ and $\gamma$ all distinct.
We then choose the strain field defined as $a_{xx}=a_0$, $a_{xy}=a_{yx}=b_0$, and the other components $a_{ij}=0$ to show that the third sum in Eq. (\ref{eq:fourth}) also cancels out: Eq. (\ref{eq:fourth}) reduces to $\varepsilon_{trial} =  (E_0 \phi/2) \left(a_0^2 \<{\e{x}}^4\> + 4 b_0^2 \<{\e{x}}^2{\e{y}}^2\> + 4 a_0 b_0 \<{\e{x}}^3\e{y}\>\right)$. Inverting the $x$ axis transforms $\e{x}$ into $-\e{x}$. As the energy must remain unchanged, one must have $\< \e{x}^3\e{y}\> =0$. 
Performing the same procedure for any pair of axes $\alpha$ and $\beta \neq \alpha$ eventually leads to the conditions $\< \e{\alf}^3\e{\bet} \> =0$.

We now choose the strain field defined as $a_{\alf\bet}=a_0 \delta_{\alf x}\delta_{\bet x}$ (this is a particular case of the field defined above). Eq. (\ref{eq:fourth}) then reduces to $\varepsilon_{trial} =  (E_0 \phi/2) a_0^2 \<{\e{x}}^4\>$. Since $\varepsilon_{trial}$ must remain unchanged by permutation of the axes, the quantity $\mcal{T}=\<{\e{\alpha}}^4\>$ must be independent of the $\alpha$ axis.
Similarly, using the strain field $a_{\alf\bet}=a_0 \delta_{\alf x}\delta_{\bet y}$, it comes that the quantity $\mcal{S}=\< (\e{\alf}\e{\bet})^2 \> $ is the same for all $\alf$ and $\bet\neq\alf$.

Using the identity $\sum_{\alpha=1}^{d} {e_{ij}^{\alpha}}^2=1 $, a first relation between $\mcal{T}$ and $\mcal{S}$ is easily obtained: $d\mcal{T}+d(d-1)\mcal{S}=1$. We obtain a second relation by using the invariance by rotation of the axes. For instance, a 45$^\circ$ rotation around the $z$ axis ``transforms'' $\e{x}$ into $(\e{x}+\e{y})/\sqrt{2}$. The equality $\mcal{T}=\e{x}^4=(\e{x}+\e{y})^4/4$ then leads to $\mcal{T}=3\mcal{S}$.
Thus, $\mcal{S}=\mcal{T}/3=1/(d(d-1))$. Finally, rearranging the terms in (\ref{eq:fourth}) leads to the expression (\ref{eq:trialenergy}).


\begin{thebibliography}{99}                                                                                               %

\bibitem {Thorpe}M.\ F.\ Thorpe, \textit{Phys. Biol.} \textbf{4} 60--63 (2007).

\bibitem {Head}D.\ A.\ Head, A.\ J.\ Levine, and F.\ C.\ MacKintosh,
\textit{Phys.\ Rev.\ Lett.} \textbf{91}, 108102 (2003).


\bibitem {Wilhelm}J. Wilhelm and E.\ Frey, \textit{Phys.\ Rev.\ Lett.}
\textbf{91}, 108103 (2003).

\bibitem {Heussinger}C.\ Heussinger and E.\ Frey, \textit{Phys.\ Rev.\ Lett.} \textbf{96}, 017802 (2006).


\bibitem {Buxton}G.\ A.\ Buxton and N.\ Clarke, \textit{Phys.\ Rev.\ Lett.} \textbf{98}, 238103 (2007).


\bibitem {Gibson}L. J. Gibson and M. F. Ashby, Cellular Solids - Structure and properties, Cambridge Univ. Press (1997, 2$^{nd}$ edition).

\bibitem {Ashby}M.F.\ Ashby, \textit{Philosophical transactions of the Royal Society A} \textbf{364}, 15-30 (2006).

\bibitem {Deshpande}V.\ S. Deshpande, M. F. Ashby and N.\ A.\ Fleck,
\textit{Acta mater.} \textbf{49}, 1035-1040 (2001).

\bibitem {Roberts}A. P. Roberts and E. J. Garboczi, \textit{J. Mech. Phys.
Solids} \textbf{50}, 33-55 (2002).

\bibitem {Christensen}R. M. Christensen, \textit{J. Mech. Phys. Sol.}
\textbf{34}, 563-578 (1986).


\bibitem {Landau}L.\ Landau and E.\ Lifchitz, Theory of elasticity, Pergamon Press, New York (1986).

\bibitem{DiDonna} B. A. DiDonna and T. C. Lubensky, \textit{Physical Review E} \textbf{72}, 066619 (2005).

\bibitem{umfpack}T. A. Davis, \textit{ACM Transactions on Mathematical Software}, \textbf{30}, no. 2, p. 196-199 (2004).

\bibitem{noteaboutM}The periodic boundary conditions impose a zero lateral strain under uniaxial load. That is why the simulations under uniaxial load gives the longitudinal modulus $M$ instead of Young's modulus $E$.

\bibitem {Liu}J. Liu, G. H. Koenderink, K. E. Kasza, F. C. MacKintosh, and D. A. Weitz, \textit{Phys. Rev. Lett.} \textbf{98}, 198304 (2007).

\bibitem {Dunlop}J. Dunlop, W. Richard, P. Fratzl and Y. Br\'{e}chet, \url{http://hdl.handle.net/2042/15749}.

\bibitem {Thorpe3}D. J. Jacobs and M. F. Thorpe, \textit{Phys. Rev. E}
\textbf{53}, 3682-3693 (1996).

\bibitem {Heussinger2}C. Heussinger, B. Schaefer, and E.\ Frey, \textit{Phys. Rev. E} \textbf{76}, 031906 (2007).

\bibitem {Kellomaki} M. Kellom\"{a}ki, J. \AA str\"{o}m, and J. Timonen
 \textit{Phys. Rev. Lett.} \textbf{77}, 2730-2733 (1996).
 
\bibitem{note42}Since the network deforms primarily though the beam stretching mode, the displacement of point $\mathbf{r}$ belonging to beam $(i,j)$ is $\mathbf{u}_{ij}(\mathbf{r})=\mathbf{A}_{ij}\centerdot\mathbf{r}$ (piecewise affine function).

\bibitem{Gurtner-Durand1}G. Gurtner and M. Durand, \textit{EPL} \textbf{87}, 24001 (2009).

\bibitem{note1}It can be noticed that no assumption is made on the relative importance of $\kappa_n$ and $\kappa_{b}$.

\bibitem {Hashin-Shtrikman2d}Z.\ Hashin, \textit{J. Mech. Phys. Solids}
\textbf{13}, 119-134 (1965).

\bibitem {Torquato}S.\ Torquato, L.\ V. Gibiansky, M.\ J. Silva and L.\ J.
Gibson, \textit{Int. J. Mech. Sci.} \textbf{40}, 71-82 (1998).

\bibitem {Hashin-Shtrikman}Z.\ Hashin and S.\ Shtrikman, \textit{J. Mech.
Phys. Solids} \textbf{11}, 127-140 (1963).

\bibitem {Durand1}M.\ Durand, J.-F. Sadoc, and D.\ Weaire, \textit{Proc. R. Soc. Lond. A} \textbf{460}, 1269-1285 (2004).

\bibitem {Durand2}M.\ Durand and D.\ Weaire, \textit{Phys. Rev. E}
\textbf{70}, 046125 (2004).

\bibitem{Hutchinson1} R. G. Hutchinson and N. A. Fleck, Microarchitectured cellular solids—the hunt for statically determinate periodic trusses. \textit{ZAMM} \textbf{85}, 607–617 (2005). 

\bibitem{Hutchinson2} R. G. Hutchinson and N. A. Fleck, The structural performance of the periodic truss, \textit{J. Mech. Phys. Solids} \textbf{54}, 756–782 (2006).

\bibitem{Sun} K. Sun, A. Souslov, X. Mao, and T. C. Lubensky, Surface phonons, elastic responsen and conformal invariance in twisted kagome lattices, \textit{PNAS} \textbf{31}, 109, 12369-12374 (2012).


\bibitem{Gurtner-Durand3}G. Gurtner and M. Durand, to be published.



\bibitem {note4}It is worth mentioning that this lattice has optimal conductivity for any value of the cross-section ratio.

\bibitem{Weaire} D. Weaire (ed) The Kelvin Problem: Foam Structures of Minimal Surface Area (Taylor and Francis, London), 1996.


\end{thebibliography}
\end{document}